\documentclass[12pt]{article}

\usepackage[utf8]{inputenc}
\usepackage[english]{babel}

\usepackage{natbib}
\usepackage[colorlinks=true,linkcolor=blue,citecolor=blue]{hyperref}
\usepackage{algorithm}
\usepackage{algpseudocode}
\usepackage{url}            % simple URL typesetting
\usepackage{booktabs}       % professional-quality tables
\usepackage{nicefrac}       % compact symbols for 1/2, etc.
\usepackage{microtype}      % microtypography
\usepackage{xcolor}         % colors

\usepackage{amssymb,amsmath,amsfonts,eurosym,geometry,ulem,graphicx,caption,color,setspace,sectsty,comment,footmisc,pdflscape,array,hyperref,subcaption,amsthm}
\newtheorem{theorem}{Theorem}
\newtheorem{corollary}[theorem]{Corollary}
\newtheorem{proposition}{Proposition}
\newtheorem{lemma}{Lemma}
\theoremstyle{remark}
\newtheorem{remark}{Remark}
\newtheorem{assumption}{Assumption}
\theoremstyle{definition}
\newtheorem{definition}{Definition}

\newtheorem*{notation}{Notation}

\newtheoremstyle{noparentheses}
  {\topsep}   % ABOVESPACE
  {\topsep}   % BELOWSPACE
  {\itshape}  % BODYFONT
  {0pt}       % INDENT (empty value is the same as 0pt)
  {\bfseries} % HEADFONT
  {.}         % HEADPUNCT
  {5pt plus 1pt minus 1pt} % HEADSPACE
  {\thmname{#1} \thmnumber{#2} \thmnote{#3}}   
\theoremstyle{noparentheses}
\newtheorem{theorem*}[theorem]{Theorem}
\newtheorem*{repeattheorem}{Theorem}
\newtheorem*{repeatlemma}{Lemma}
\newtheorem*{repeatproposition}{Proposition}
\newtheorem*{repeatcorollary}{Corollary}

\usepackage{amsfonts}

\def\*{\star}

\usepackage{graphicx}
\usepackage{mathtools}
\usepackage{footnote}
\usepackage{float}
\usepackage{xspace}
\usepackage{multirow}
\usepackage{wrapfig}
\usepackage{framed}
\usepackage{xcolor}

\newcommand{\field}[1]{\mathbb{#1}}

\newcommand{\fR}{\field{R}}

\newcommand{\calZ}{{\mathcal{Z}}}

\newcommand{\calC}{{\mathcal{C}}}

\newcommand{\one}{\boldsymbol{1}}

\begin{document}

\begin{titlepage}

\title{Causal Bandits: Online Decision-Making in Endogenous Settings}
\date{}
\author{
  Jingwen Zhang\footnote{Jingwen Zhang is a third-year Ph.D. student at Foster School of Business, University of Washington,  \texttt{jingwenz@uw.edu}.} \and
   Yifang Chen\footnote{ Yifang Chen is a third-year Ph.D. student at Paul G. Allen School of Computer Science and Engineering, University of Washington, \texttt{yifangc@cs.washington.edu}.} \and 
   Amandeep Singh\footnote{Amandeep Singh is an Assistant Professor at Foster School of Business, University of Washington,  \texttt{amdeep@uw.edu}.} 
 }
\maketitle

\begin{abstract}
  The deployment of Multi-Armed Bandits (MAB) has become commonplace in many economic applications. However, regret guarantees for even state-of-the-art linear bandit algorithms (such as Optimism in the Face of Uncertainty Linear bandit (OFUL)) make strong exogeneity assumptions w.r.t. arm covariates, i.e., the covariates are uncorrelated with the random error. This assumption is very often violated in many economic contexts and using such algorithms can lead to suboptimal decisions. Further, in social science analysis, it is also important to understand the asymptotic distribution of estimated parameters. To this end, in this paper, we consider the problem of online learning in linear stochastic contextual bandit problems with endogenous covariates. We propose an algorithm we term $\epsilon$-\textit{BanditIV}, that uses instrumental variables to correct for this bias, and prove an $\tilde{\mathcal{O}}(k\sqrt{T})$ upper bound for the expected regret of the algorithm, where $k$ is the dimension of the instrumental variable and $T$ is the number of rounds in the algorithm. Further, we demonstrate the asymptotic consistency and normality of the $\epsilon$-\textit{BanditIV} estimator. We carry out extensive Monte Carlo simulations to demonstrate the performance of our algorithms compared to other methods. We show that $\epsilon$-\textit{BanditIV} significantly outperforms other existing methods in endogenous settings. Finally, we use data from real-time bidding (RTB) system to demonstrate how $\epsilon$-\textit{BanditIV} can be used to estimate the causal impact of advertising in such settings and compare its performance with other existing methods.\\
% \noindent\textbf{Keywords:} Multi-Armed Bandits, Causal Inference, Online Learning, Instrumental Variables\\

\bigskip
\end{abstract}
\setcounter{page}{0}
\thispagestyle{empty}
\end{titlepage}

\section{Introduction}

The proliferation of user-level data presents a unique challenge in front of decision-makers. Decision-makers want to use individual-level data to tailor their decisions for each user. However, given the dynamic nature of online platforms, decision-makers need to adopt these decisions with incrementally available data. \citet{woodroofe1979one} first proposed a simple model to solve such sequential decision-making problems with covariates. \citet{langford2007epoch} later named the model "contextual bandit". Contextual linear bandits have been adopted across  a wide variety of applications from advertising \citep{tang2015personalized, aramayo2022multi}
to healthcare \citep{durand2018contextual}, dialogue systems \citep{liu2018customized}, and personalized product recommendations \citep{li2010contextual,qin2014contextual}. In the setting of a contextual linear bandit problem, in each round, the decision-maker observes a set of actions with each action characterized by a set of features. Decision-maker selects an action and observes a reward corresponding to that action. The objective of the decision-maker is to achieve cumulative reward close to that of optimal policy in hindsight.

\begin{figure}
    \centering
    \includegraphics[scale=0.45]{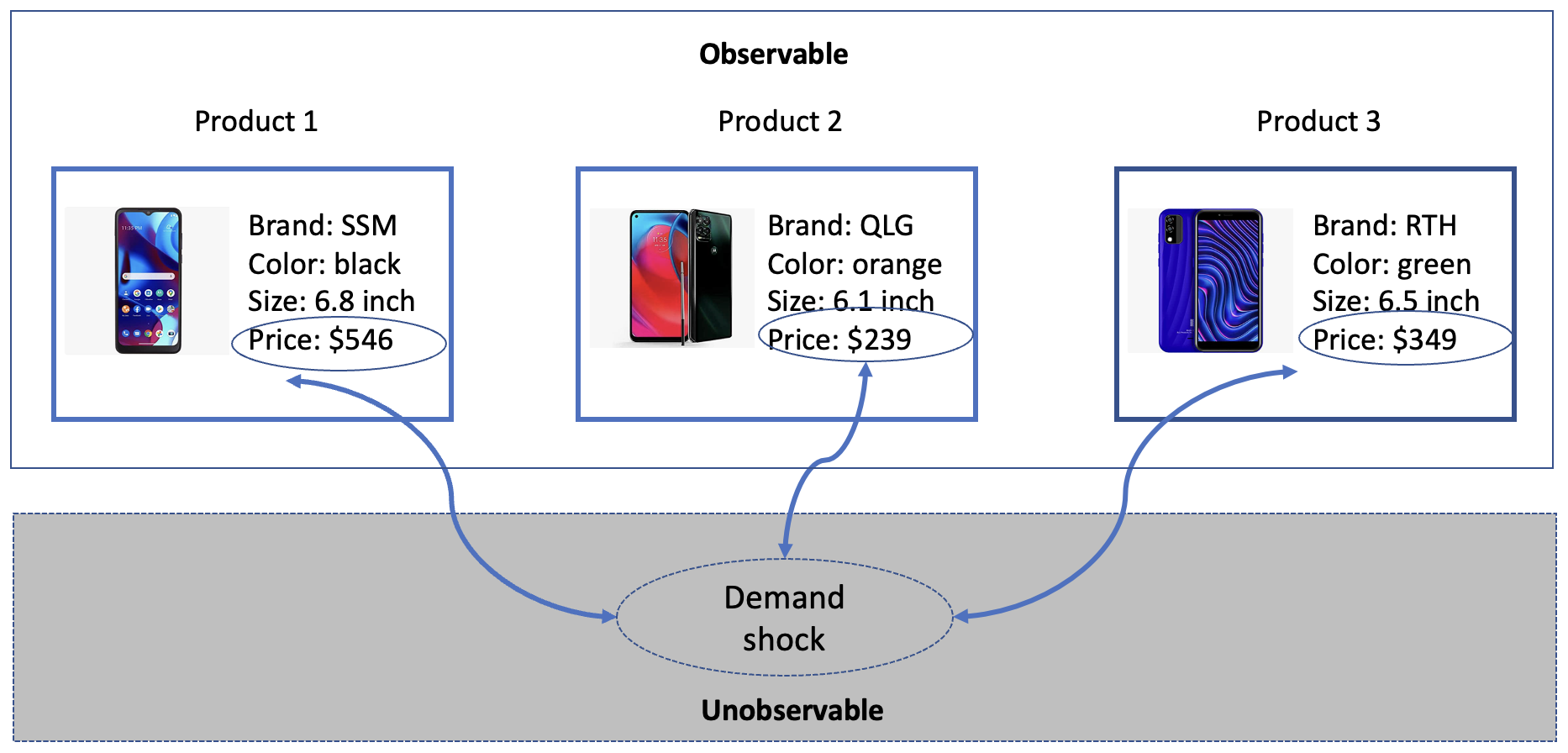}
    \caption{Illustration for endogeneity in product recommendation}
    \label{fig:end_ill}
\end{figure}

Traditional formulations of contextual bandits make the unconfoundedness assumption i.e., the arm covariates are exogenous and are uncorrelated with the unobserved noise. However, in many economic settings, arm features can be correlated with unobserved noise. For instance, consider the problem of generating product recommendations for consumers. Platform operators would usually run online experiments to uncover the relationship between product features and consumer demand. Generally in such settings, product observable features like price are controlled by product owners (different from the platform operator) which could be set in anticipation of consumer demand and hence be correlated with the demand shocks unobserved by the platform operator (see Figure \ref{fig:end_ill} for illustration). In such settings, traditional bandit algorithms might lead to sub-optimal decisions. A common approach to correct the endogeneity bias in the offline setting is to use instrumental variables. Instruments
 are a set of variables correlated with endogenous variables but are otherwise not associated with
the outcome variable. Offline instrumental variable methods use the variation in the exogenous component of the endogenous variable induced by the
variation in the instrumental variables to make inference of causal effects. In this work, we propose an online instrumental variable method, we term $\epsilon$-\textit{BanditIV} to address the issues induced by endogenous features in online settings and estimate the relationship between rewards and features.

Next, literature in contextual linear bandit has also primarily focused on minimizing the expected regret of the algorithm \citep{auer2002using, dani2008stochastic, chu2011contextual, abbasi2011improved}. However, in a variety of social science problems, one may also be interested in understanding the asymptotic distribution of the estimated parameters. Algorithms that just minimize the regret may not guarantee the consistency of the model parameters and might lead to non-standard asymptotic distribution. Further, having standard asymptotic distributions, one can easily test the significance of features, and
 offers a traditional way to select variables. For instance, in the product recommendation example, an online platform might also be interested in understanding the effect of various marketing-mix variables (like price, and promotion) on consumer demand. To this end, we propose a linear bandit algorithm that pursues both regret minimization and consistent estimation of model parameters in endogenous settings. We use the martingale central limit theorem to show that our estimator of model parameters is asymptotically normal. 

Finally, we demonstrate how $\epsilon$-\textit{BanditIV} algorithm can be used to place optimal bids and simultaneously measure the average treatment effect (ATE) of advertising in Real-time bidding (RTB) auctions. In targeted marketing, RTB has become a dominating mechanism for selling user exposure to  advertisers. Real-time bidding (RTB) systems, utilize auctions to allocate user impressions to
competing advertisers. Even after their widespread adoption, measuring the effectiveness of advertising in such systems has remained hard in practice. Recent literature \citep{johnson2017ghost} has characterized the inherent endogeneity in such settings and challenges in designing scalable experimental designs. On a high level, there are two key challenges that one needs to account for while designing experiments for RTB systems. The first challenge is that estimating the ATE of advertising requires randomly exposing users to advertising and then comparing the outcomes of users exposed to advertising to those who were not. However, in RTB settings, exposures are strictly not in the control of the bidder or experimenter and are determined through an auction. 

The second challenge is the cost of the experimentation. To induce randomized exposure across the customer populations, the experimenter has to intentionally win or lose auctions. When bidding is not optimized, experimentation could lead to significant costs. For instance, in absence of optimal bidding function, overbidding to win an auction could lead to higher ad costs. Similarly, unnecessarily losing auctions could lead to lower consumer exposure and hence revenue. To this end, demonstrate how $\epsilon$-\textit{BanditIV} can be used to tackle both these challenges. We also propose examples of plausible exogenous data that can be used as instrumental variables in such settings.       We finally compare our method with existing online algorithms including  Optimism in the Face of Uncertainty Linear bandit (OFUL) 
and Thompson Sampling (TS). We find significant advantages of our algorithm in both expected regret and parameter inference.

To summarize, our paper makes the following contributions --
\begin{itemize}
\item We study the stochastic linear bandit problem with endogenous features. We propose a new algorithm, we term $\epsilon$-\textit{BanditIV}, which incorporates instrumental variables to correct for the bias induced by endogenous features and show that the total expected regret of the algorithm is upper bounded by $\tilde{\mathcal{O}}(k\sqrt{T})$. 
\item Next, as in many economic contexts, researchers might not only be interested in minimizing the regret over the outcomes but also in conducting inference over estimated parameters. To this end, we establish the asymptotic consistency and normality of the estimator.
   
    \item Finally, we demonstrate the application of $\epsilon$-\textit{BanditIV} algorithm to real-time bidding systems. Conducting causal measurements of ad exposures while minimizing the cost of experimentation has remained hard for advertisers in the RTB settings. Through application to real-world data, we show  $\epsilon$-\textit{BanditIV} can effectively measure the ATE in such settings while maintaining close to oracle regret.  
\end{itemize}
The rest of the paper is organized as follows. We review related literature in areas of marketing, computer science, and econometrics in Section \ref{sec: literature}. We set up the main problem, i.e. endogeneity problem in online stochastic linear bandits in Section \ref{sec: problem}. We formulate the $\epsilon$-\textit{BanditIV} algorithm in Section \ref{sec: algorithm} and show regret bound in Section \ref{sec: regret}. We derive theoretical analysis on consistency and normality of the estimator used in the $\epsilon$-\textit{BanditIV} algorithm in Section \ref{sec: inference}. We conduct simulations of the proposed algorithm on synthetic data and RTB data in Sections \ref{sec: simulation} and \ref{sec: rtb} respectively. 

\begin{notation}
Throughout this paper, we use $||\cdot||_p$ to denote the $p$-norm of a vector or a matrix. We use $||\cdot||_F$ to denote the Frobenius norm for a matrix. For a vector $x$ and positive definite matrix $A$, we denote $\sqrt{x'Ax}$ as $||x||_A$. We use $\langle \cdot, \cdot \rangle$ as the inner product. We denote the sequence $a_0, a_1, \cdots, a_{\infty}$ as $\{a_t\}_{t=0}^{\infty}$. For a matrix $X$, the $i^{th}$ column is denoted as $X^{(i)}$ and the $j^{th}$ element of the $i^{th}$ column is denoted as  $X^{(i)(j)}$.
\end{notation}

\section{Literature review}\label{sec: literature}
Our paper contributes to three different strands of literature  (i) adaptive experimentation and digital advertising, (ii) interactive learning and inference with MABs, and (iii) machine learning and econometrics.

% \subsection{Digital Experimentation and Advertising (Targeted Marketing)}
% the following paragraph is hidden for a shorter version (\le 6000 words) on 11/21/2022
% Recently, there has been a surge of papers studying adaptive experimentation, recommendation, and targeted advertising.
The unprecedented growth of digital advertising markets attract increasing attention from both industry and academia. To improve the allocative efficiency of digital advertising market, there has been a surge of papers studying the advertising effect. The advertising effect is the incremental number of outcomes obtained as a result of an advertising campaign, such that these outcomes would not have happened without the campaign \citep{gordon2021inefficiencies}. \cite{zantedeschi2017measuring} proposed a hierarchical Bayesian model for advertising effects considering individual differences in purchase propensity and marketing response. \cite{rafieian2021targeting} developed a unified modeling framework to answer a set of questions which includes the advertising effect of adopting efficient targeting policies. 
% \cite{rafieian2022variety} examined the advertising effect of an increase in the ad variety on the next ad. 
However, a number of challenges regarding the measurement of advertising effects remain. \cite{gordon2019comparison} empirically compared multiple observational models for causal effects of digital advertising and suggested that common observational approaches fail to accurately measure the true causal effect. \cite{gordon2021inefficiencies} pointed out several problems with experiments on ad effects, such as large sample size required and high experimental costs. 

Adaptive experiments are increasingly used to investigate the advertising effect. \cite{araman2022diffusion} studied a Bayesian sequential experimentation problem through dynamic programming and diffusion-asymptotic analysis. \cite{misra2019dynamic} proposed a dynamic price experimentation policy which extended MAB algorithms to include microeconomic choice theory. \cite{gur2022adaptive} considered auxiliary information between decision epochs into the development of an adaptive policy to improve experimentation performance. \cite{bertsimas2019covariate} proposed an online allocation algorithm in experimental clinical trials using robust mixed-integer optimization to achieve a high statistical power. \cite{moazeni2020sequential} modeled the marketing campaign performance as a multiplicative advertising exposure problem and proposed a computationally efficient learning policy through solving sequentially mixed-integer linear optimization problems. \cite{delshad2022adaptive} studied the personalized dose-finding clinical trial problem through a Bayesian framework by stochastic programming and proposed an adapted one-step look-ahead approximate policy. \cite{anderer2022adaptive} introduced a Bayesian adaptive clinical trial design that combined both surrogate and true outcomes to improve trial performance. 
% In order to optimize the reward from experimentation, it if of great importance that we learn the causal effect of a treatment accurately. As a result, a stream of literature in online advertising focuses on how to estimate the causal effect of advertisements. 
% Adaptive experimentation and dynamic recommendation for online display are two common applications of bandits in marketing contexts. There has been extensive literature contributing to digital experimentation and advertising. 
% There has been a a surge of papers in advertising focusing on the improvement of estimation for causal effects of advertisements.
% \cite{amaldoss2015keyword} studied first-page bid estimate (FPBE) mechanism proposed by Google in keyword search advertising and demonstrated the advantages of the FPBE over the generalized second-price auction, which was typically used when search engine sells ad slots. 
% \cite{amaldoss2016keyword} examined the strategic role of management costs and broad match mechanism in sponsored keyword search advertising. 
\cite{bumbaca2020scalable} proposed an algorithm to efficiently estimate Bayesian hierarchical models for inferences when researchers have limited number of observations per consumer in online targeted marketing. 
% \cite{choi2020online} reviewed the display advertising literature among both theoretical and empirical research. 
%\cite{bertsimas2007learning} proposed a Bayesian formulation of an online decision problem in interactive marketing contexts in an early work. They provided heuristics and simulated results to show their methods outperform previous approaches. 

At the framework level, several studies formulated adaptive experimentation and targeted advertising problems as MAB. \cite{schwartz2017customer} utilized MAB to solve the online ad allocation problem. \cite{aramayo2022multi} studied complete collections of ads and nonstationary rewards in house ads recommendation problem by using a MAB scheme. 
% \cite{li2010contextual} applied a contextual bandit framework to model personalized news article recommendation. 
\cite{tang2015personalized} presented implementations of contextual bandit algorithms on 
% a news recommendation problem and 
an online advertising problem. \cite{aramayo2022multi} studied a contextual bandit approach to dynamically exhibit house ads to customers.
% Especially, this paper is closely related with CMAB approach, which is a popular branch of models in MAB. 
% \subsection{Interactive Learning and Inference with MABs}
We focus on the sequential decision making under the linear contextual bandit framework. \cite{auer2002using} first introduced the contextual bandit setting  through the Linear Reinforcement Learning (LinRel) algorithm with linear value functions. Subsequently, the contextual framework was improved by \cite{dani2008stochastic}, \cite{chu2011contextual} and \cite{abbasi2011improved} through Upper Confidence Bound (UCB) type algorithms.  
% further modified the analysis of the linear bandit problem and improved the regret bound by a logarithmic factor. 
% More recently, 
% the following sentence is hidden for a shorter version (\le 6000 words) on 11/21/2022
% \cite{bastani2021mostly} studied performance of exploration-free policies in contextual bandit settings and proposed the Greedy-First algorithm which only utilizes observed contexts and rewards to decide whether to follow a greedy algorithm or to explore. 
This stream of literature often requires independency between contexts and the random error term, which however cannot be always satisfied in the reality. We consider a setting which allows the dependency between contexts and error term. This more general setting is well suited to real-world applications where endogeneity problems happen.

% In online settings, data arrives sequentially. 
However, machine learning methods, such as methods used in MAB problems, are usually designed to obtain the best out-of-sample predictions, while econometric methods are focused on deriving best unbiased estimators \citep{dzyabura2018machine}. In this paper, we are interested in both reaching good predictions and unbiased causal estimation. Yet, the adaptive nature of MAB problem complicates the causal inference of advertising effects. \cite{nie2018adaptively} showed that estimates of arm-specific expected rewards in UCB and TS algorithms are biased downwards. This downward bias is due to that arms with random upward fluctuation are sampled more, while those arms with downward fluctuation are sampled less (\cite{hadad2021confidence}). \cite{bibaut2021post} presented that standard estimators no longer follow normal distribution asymptotically so that classic confidence intervals fail to provide correct coverage.
To address this inference problem in online settings, \cite{chen2021statistical} studied the consistency and asymptotic distribution of the online ordinary least square estimator under epsilon-greedy policy. We extend their approach to two stage least square estimation in online settings and contribute to the literature by providing the consistency and asymptotic normality of our estimator. Besides, \cite{lattimore2016causal} studied the bandit problem on a causal graph and proposed new algorithms to learn good interventions in stochastic online settings. There are two significant differences between their work and ours. First, we focus on a single linear model of the features and rewards, while they studied causal graph of features and rewards. Second, they did not consider the endogeneity problem, which our work mainly contributes to.

% \subsection{Machine learning and Econometerics}

This paper is also closely related to the stream of literature in instrumental variable methods for endogeneity problems. \cite{griliches1977estimating}, \cite{hausman1983specification}, \cite{angrist1995identification}, and \cite{chen2005measurement} provided theoretical analysis of instrumental variables in linear models.
% while \cite{hansen1982large}, \cite{ai2003efficient}, \cite{newey2003instrumental} and \cite{chernozhukov2007instrumental} studied instrumental variables in nonlinear models. 
% \cite{imbens2014instrumental} reviewed work on instrumental variable methods and discussed applications and restrictions. 
Different from the standard data analysis framework in offline settings, the endogeneity problem can be exacerbated during the dynamic interaction between data generation and data analysis in online settings \citep{li2021causal}. This paper builds on existing instrumental variable methods in offline settings and considers the dynamic interaction by utilizing the CMAB framework. For these reasons, the paper contributes to the literature by combining both CMAB models and instrumental variable methods to improve the estimation of causal effect of ad exposure and hence achieve better performance of decision policies.

Lastly, this paper is related with the literature in RTB. \cite{choi2020online} reviewed research work in RTB questions regarding both advertisers, like how to calculate the impression value, and publishers, like how to allocate impressions across advertisers. \cite{sayedi2018real} used a game theory framework to study the effects of RTB on publisher' revenue, advertisers' profits as well as their strategies. \cite{balseiro2022dynamic} studied an auction mechanism design to satisfy two properties including no positive transfers and periodic individual rationality, which can reach asymptotically first best for the platform. \cite{balseiro2015repeated} examined the equilibrium in repeated auctions considering strategic responses of budget-constrained bidders and provided implications on auction mechanism design for publishers. 
In this paper, instead of addressing questions for publishers, we focus on the questions advertisers face in RTB. We aim to learn the optimal bidding price and the effect of an ad exposure through the algorithms we propose.

\section{Problem Setting}\label{sec: problem}
Let $T$ be the number of rounds and $K$ the number of arms in each round. In each round $t$, the learner observes a feature vector $x_{t,a}\in \mathbb{R}^d$, $a\in \{1,\cdots,K\}$, with $||x_{t,a}||_2\le L_x$, for each arm. We use the subscript $a$ to represent the $a^{th}$ arm from the $K$-arm set and $L_x$ to represent a constant upper bound for $||x_{t,a}||_2$. After observing the feature vectors, the learner selects an arm $a_t$ and receives a reward $y_{t,a}\in \mathbb{R}$, with $|y_{t,a}|\le L_y$ where $L_y$ is a constant upper bound for $|y_{t,a}|$. Under linear realizability assumption, we have $\mathbb{E}[y_{t,a}|x_{t,a}]=x'_{t,a}\beta_0$ for all $t$ and $a$, where $\beta_0\in \mathbb{R}^d$ is an unknown true coefficient vector. We specifically assume $y_{t,a}$ as the following.
\begin{equation}\label{eq: y_ta}
    y_{t,a} = \beta'_0x_{t,a} + e_{t,a}
\end{equation}
where $e_{t,a}\in \mathbb{R}$ is a 1-subgaussian error term, s.t. $\mathbb{E}[e_{t,a}]=0$. We denote the cumulative distribution of  $e_{t,a}$ as $\mathcal{P}_e$. Unlike traditional bandit algorithms, we have a twofold goal in this problem: (i) to estimate the main coefficient of interest $\beta_0$ (ii) to achieve the largest reward by selecting the optimal arm.

Standard Contextual Linear Bandit settings have $\mathbb{E}[e_{t,a}x_{t,a}]=0$, which implies that the error term is independent with the feature vector. However, this assumption can be violated in various cases such as when we have omitted variable in the error term, measurement error in the regressor, simultaneous equation estimation, and etc. Mathematically, when we have the following,
\begin{equation*}
    \mathbb{E}[e_{t,a}x_{t,a}]\neq0
\end{equation*}
the endogeneity problem occurs. 
Ordinary Least Square (OLS) estimator for the coefficient of interest $\beta_0$ is inconsistent in endogenous settings. 
Econometrics literature uses Instrumental Variable (IV) method to address the endogeneity problem. A variable $z_{t,a}$ is a valid instrumental variable if it satisfies the following three conditions \citep{m2014introductory}: (i) it is uncorrelated with the error term $e_{t,a}$, i.e. $\mathbb{E}[z_{t,a}e_{t,a}]=0$, (ii) it is correlated with the endogenous covariates $x_{t,a}$, (iii) it has no direct effect on the reward $y_{t,a}$. Consider a valid instrumental variable $z_{t,a}\in \mathbb{R}^k$, with $||z_{t,a}||_2\le L_z$ where $L_z$ is a constant upper bound for $||z_{t,a}||_2$. We assume $\mathbb{E}[z_{t,a}x'_{t,a}]\in \mathbb{R}^{k\times d}$ has full column rank $d$.\footnote{The assumption $\mathbb{E}[z_{t,a}x'_{t,a}]\in \mathbb{R}^{k\times d}$ has full column rank $d$ implies that $k\ge d$.} and the minimum eigenvalue of $\mathbb{E}[z_{t,a}z_{t,a}']$ is positive.
\begin{equation}\label{eq: x_ta}
    x_{t,a} = \Gamma'_0z_{t,a}+u_{t,a}
\end{equation}
where $\Gamma_0\in \mathbb{R}^{k\times d}$ is an unknown true coefficient vector, each element in $u_{t,a}\in \mathbb{R}^{d}$ is a 1-subgaussian error term with mean zero and the independency condition $\mathbb{E}[z_{t,a}u'_{t,a}]=0$ is satisfied by construction. Notice that by plugging the equation of $x_{t,a}$ (Equation (\ref{eq: x_ta})) into Equation (\ref{eq: y_ta}), we have
\begin{align}
    y_{t,a}=(\Gamma_0\beta_0)'z_{t,a}+v_{t,a}
\end{align}
where $v_{t,a}=\beta'_0u_{t,a}+e_{t,a}$. We can prove that each element of $v_{t,a}$ is $(||\beta_0||_2^2 +1)$-subgaussian (we provide the proof in Appendix B). We denote $\Gamma_0\beta_0$ as $\delta_0$ for the simplicity of analysis.

Two-Stage Least Squares (TSLS) is an instrumental variable method commonly used in offline settings to correct estimation bias in the endogeneity problem. Consider vector $Y\in \mathbb{R}^T$ with $y_{t,a}$ as its elements, matrix $X\in \mathbb{R}^{T\times d}$ with $x'_{t,a}$ as its rows, matrix $Z\in \mathbb{R}^{T\times k}$ with $z'_{t,a}$ as its rows, where $t\in \{1,\cdots, T\}$. We briefly illustrate TSLS estimation procedure in offline settings as the following: (i) First, regress the set of feature vectors $X$ on the set of instrumental variable vectors $Z$ using OLS method to obtain an estimated sample features $\hat{X}$; (ii) Then regress $Y$ on the estimated feature vectors $\hat{X}$ to obtain the estimator $\hat{\beta}_{\hat{X},Y}$ using OLS method again. 
\begin{definition}
    (Two-Stage Least Squares Estimator) 
    \begin{align*}
        \hat{X} = P_Z X, P_Z = Z(Z'Z)^{-1}Z'\\
        \hat{\beta}_{\hat{X},Y} = (\hat{X}'\hat{X})^{-1}\hat{X}'Y 
    \end{align*}
\end{definition}

% {\color{red}shouldnt this be $\hat{\beta}_{\hat{X},Y}$}\jw{corrected}
To facilitate further analysis on the TSLS estimator, we also define related OLS estimators as the following, where $\hat{\delta} = \hat{\Gamma} \hat{\beta}_{\hat{X},Y}$ (to see this equation, refer Appendix B for the proof).
\begin{definition}
    (First Stage OLS Estimator)
    \begin{align*}
        \hat{\Gamma} = (Z'Z)^{-1}Z'X
    \end{align*}
\end{definition}
\begin{definition}
    (OLS Estimator Based on Instrumental Variable)
    \begin{align*}
        \hat{\delta} = (Z'Z)^{-1}Z'Y
    \end{align*}
\end{definition}

It is known that TSLS estimator is consistent in offline settings. In this paper, we adopt the standard two-stage least square procedure to the online setting and demonstrate how one can use instrumental variables to address the issue of endogeneity in linear contextual bandits. 

We aim to design an online decision-making algorithm that learns the coefficient of main interest $\beta_0$ so that we can maximize the total expected reward after pulling arms under the endogeneity problem. We define the total expected regret of an algorithm after $T$ rounds as 
\begin{align*}
    R_T=\sum_{t=1}^T\mathbb{E}[y_{t,a_t^*} - y_{t,a_t}]
    =\sum_{t=1}^T\mathbb{E}[ \langle \Gamma'_0z_{t,a_t^*}, \beta_0 \rangle- \langle  \Gamma'_0z_{t,a_t} , \beta_0 \rangle]
\end{align*}
where $a_t^*=\arg\max_{a}  \langle \Gamma'_0z_{t,a}, \beta_0 \rangle$ is the best arm at round $t$ according to the true coefficient vectors $\beta_0$ and $\Gamma_0$, and $a_t$ is the arm selected by the algorithm at round $t$.

We relist key assumptions in this paper as the following,

\begin{assumption}\label{asm: l2norm}
    $||z_{t,a}||_2\le L_z$, $||x_{t,a}||_2\le L_x$, $||y_{t,a}||_2\le L_y$ for all $t$ and $a$.
\end{assumption}
\begin{assumption}\label{asm: mineig}
    The minimum eigenvalue of $\mathbb{E}[z_{t,a}z_{t,a}']$ is larger than a positive constant $\lambda$.
\end{assumption}

% {\color{red} Do not use the term "l2 norm" or "infintiy norm" -- just use the technical notation}

Assumption \ref{asm: l2norm} ensures that the $||\cdot||_2$ of instrumental variables, feature vectors, and the reward are bounded by positive constants. Notice that the $||\cdot||_{\infty}$ of a vector is smaller than the $||\cdot||_2$ of the vector, which implies that the $||\cdot||_{\infty}$ of the instrumental variables, features , and the reward are also upper bounded by these constants, $L_z$, $L_x$, and $L_y$ respectively. 
% Meanwhile, $||z_{t,a}||_2\le L_z$ implies that the mean of the martingale differences will converge to 0 \jw{need explain what is the martingale difference}.
Assumption \ref{asm: mineig} guarantees that, with high probability, the sample second moment $\sum_{s=1}^t z_{s,a}z_{s,a}'$ is non-singular so that the OLS estimators $\hat{\Gamma}$ and $\hat{\delta}$ exist.
We need these assumptions to bound the total regret and the inference bias.

\section{BanditIV Algorithm}\label{sec: algorithm}
In this section, we propose the $\epsilon$-\textit{BanditIV} Algorithm (Algorithm \ref{algo: epsilon_banditiv}) by incorporating instrumental variables in online settings. The algorithm is based on existing linear bandit algorithms, especially the OFUL algorithm proposed by \cite{abbasi2011improved}. 

The $\epsilon$-\textit{BanditIV} algorithm takes as input initial regularization parameters $\gamma_z, \gamma_x>0$, confidence set parameters $\{G_t\}_{t=1}^{\infty}$, $\{B_t\}_{t=1}^{\infty}$, as well as a sequence of non-increasing exploration parameters $\{\epsilon_t\}_{t=1}^{\infty}$. We will specify the confidence set parameters in Section \ref{sec: regret}. 

\textbf{Estimation:} The algorithm maintains four matrices $U_t, V_t, W_t, Q_t$ to calculate estimated parameters in each round as described in the pseudo code of Algorithm \ref{algo: epsilon_banditiv}. If the current time $t$ is after the first round, the algorithm utilizes past-choice-related instrumental variables $Z_t$ and past choices $X_t$ to estimate $\Gamma_0$ through OLS and obtain an estimated $X_t$, which is $\hat{X}_t$. Then, based on $\hat{X}_t$ and past observations of rewards $Y_t$, the algorithm estimates $\beta_0$ by OLS again. We denote the estimator for $\beta_0$, which is $\hat{\beta}_t$, in the $\epsilon$-\textit{BanditIV} algorithm as the \textit{BanditIV estimator}.

\textbf{Confidence Sets:} Following the \textit{Optimisim in the Face of Uncertainty principle} (OFU), we need to maintain confidence sets for all unknown parameters in the model. As described in Section \ref{sec: problem}, we have two unknown parameters, $\Gamma_0$ and $\beta_0$. Thus, we construct two confidence sets $C_{1,s}$,  $C_{2,s}$, $s \in \{1,\cdots, T\}$ for the first and the second stage estimation respectively in each round. The idea for the confidence sets are to make the estimation optimistic with the conditions that "with high probability" the true coefficients are in the confidence sets and we can calculate the confidence sets from the past chosen arms $X_t$, related instrumental variables $Z_t$, and rewards $Y_t$.

\textbf{Execution:} In each round, we conduct a stochastic decision with probability $\epsilon_t$ for choosing a random arm and probability $1-\epsilon_t$ for choosing the estimated best arm. The estimated best arm generated by the algorithm is an arm $a_t$ which is related to an instrumental variable $z_t$ that can maximize the estimated reward jointly with a pair of optimistic estimates of two-stage coefficients in the confidence sets. To maximize the estimated reward, the algorithm chooses a pair of optimistic estimates  $(\tilde{\Gamma}_t, \tilde{\beta}_t)=\arg\max_{\Gamma \in C_{1,t}\\ \beta \in C_{2,t}}(\max_{z\in \mathcal{Z}_t} \langle \Gamma'z, \beta \rangle)$ and then chooses an arm $a_t$ such that $a_t=\arg \max_{a\in \mathcal{A}_t} \langle \tilde{\Gamma}'_tz_a, \tilde{\beta}_t \rangle$, where we denote the arm set at time $t$ as $\mathcal{A}_t$. Equivalently, the estimated best arm $a_t$, is stated in Equation (\ref{eq: iv_chosen}). After an arm is chosen, we observe the reward $y_{t,a_t}$ as stated in Equation (\ref{eq: y_ta}).
\begin{align}\label{eq: iv_chosen}
    a_t = \arg\max_{a\in \mathcal{A}_{t}}\max_{\Gamma \in C_{1,t} }\max_{\beta \in C_{2,t} }\langle \Gamma' z_a, \beta\rangle
\end{align}

Note that when $\epsilon_t=0$ for $t\in \{1,2\cdots,\infty\}$, the algorithm chooses the estimated best arm in each round. To simplify the analysis, we denote the algorithm when $\epsilon_t=0$ for $t\in \{1,2\cdots,\infty\}$ as \textit{BanditIV}.

\begin{algorithm}
\small
\caption{$\epsilon$-BanditIV}\label{algo: epsilon_banditiv}
\begin{algorithmic}
\State\textbf{Input: } $\gamma_z, \gamma_x$, $\{G_t\}_{t=1}^{\infty}$,$\{B_t\}_{t=1}^{\infty}$, $\{\epsilon_t\}_{t=1}^{\infty}$
\State Set $U_0=\gamma_z I\in \mathbb{R}^{k \times k},V_0=0\in \mathbb{R}^{k \times d},W_0=\gamma_x I \in \mathbb{R}^{d \times d},Q_0=0\in \mathbb{R}^{d \times 1}$
\State Set $C_{1,s}=\{\Gamma: ||\Gamma^{(i)}-\hat{\Gamma}_{s}^{(i)}||_{U_{s}}\le G_{s}\}, C_{2,s}=\{\beta: ||\beta-\hat{\beta}_{s}||_{W_{s}}\le B_{s}\}$, $\forall s\in \{1,2,\cdots\, T\}$, $\forall i\in \{1,2,\cdots\,d\}$
\State Nature reveals $\mathcal{Z}_{0}$. We randomly choose $z_0\in \mathcal{Z}_{0}$ and set $Z_0=z_0$
\State Nature reveals $\mathcal{X}_{0}$. From the set $\mathcal{X}_{0}$, we play $x_0$ which is related with $z_0$ and then observe the reward $y_0$. We set $X_0=x_0$ and $Y_0=y_0$
\For{$t:=1,2,\cdots,T$,}
    \State $U_t=U_{t-1}+z_{t-1}z'_{t-1}, V_t=V_{t-1}+z_{t-1}x'_{t-1}$
    \State $\hat{\Gamma}_t=U_t^{-1}V_t$, $\hat{X}_{t-1}=Z_{t-1}\hat{\Gamma}_t$
    \State $W_t=W_0+\hat{X}'_{t-1}\hat{X}_{t-1}$, $Q_t=Q_0+\hat{X}'_{t-1}Y_{t-1}$
    \State $\hat{\beta}_t=W_t^{-1}Q_t$
    \State Nature reveals $\mathcal{Z}_{t}$. With probability $\epsilon_t$, we uniformly choose a random $z_t \in \calZ_t$. With probability $1-\epsilon_t$, we choose $ z_{t}=\arg\max_{z\in \mathcal{Z}_{t}}\max_{\Gamma \in C_{1,t} }\max_{\beta \in C_{2,t} }\langle \Gamma'z, \beta\rangle $ and update $Z_{t}=[Z'_{t-1} \quad z_{t}]'$
    \State Nature reveals the set of arms $\mathcal{X}_t$. We play $x_t$ which is related with $z_t$ and observe the reward $y_t$. Update $X_t=[X'_{t-1} \quad x_t]', Y_t=[Y'_{t-1} \quad y_t]'$

\EndFor
\end{algorithmic}
\end{algorithm}

For the simplicity of notations, we omit the subscript $a$ of $z_{t,a}$, $x_{t,a}$, and $y_{t,a}$ from below. By $x_t$, we mean one arm chosen by the algorithm from the arm set $\mathcal{X}_{t}$ at time $t$; the variable $z_t$ is the instrumental variable related to $x_t$; the variable $y_t$ is the reward generated by choosing the arm $x_t$. By $x_t^*$, we mean the true optimal arm from the arm set $\mathcal{X}_{t}$ at time $t$; $z_t^*$ is the instrumental variable related to $x_t^*$.

\subsection{Regret Analysis} \label{sec: regret}
In this section, we give upper bounds on the regret of the $\epsilon$-\textit{BanditIV} algorithm as well as the \textit{BanditIV} algorithm. The proofs can be found in Appendix A. We first show an $\tilde{\mathcal{O}}(k\sqrt{T}$) upper bound for the total expected regret with parameters of confidence set by Theorem \ref{thm: total_regret} and Corollary \ref{thm: total_regret_eps}. Then by Lemma \ref{prop:Bt} and Lemma \ref{prop:Gt}, we present that the true coefficients are in the confidence sets with high probability and we also give accurate definitions for confidence set parameters.

% \begin{theorem}\label{thm: total_regret}
% The expected cumulative regret of the $\epsilon$-BanditIV at time T, with probability at least $1-\delta$, is upper-bounded by
% \begin{align*}
%     R_T &\le B_T \sqrt{2Td \log(\frac{T+d}{d})} + (\frac{2}{\gamma_x}+\|\beta_0\|_2)G_T\sqrt{2Tk \log(\frac{T+k}{k})} + 2\epsilon_0 T L_y
% \end{align*}
% where $B_T=\sqrt{\gamma_x} ||\beta_0||_2+\sqrt{2\log(\frac{2T}{\delta})+ d\log\left(\frac{5TL_x^2}{d} \right)}$,\\ 
% $G_T=\sqrt{\gamma_z} ||\Gamma_0^{(i)}||_2+\sqrt{2\log(\frac{2T}{\delta})+k \log\left(\frac{5TL_y^2}{k} \right)}$.
% \end{theorem}
\begin{theorem}\label{thm: total_regret}
The expected cumulative regret of the $\epsilon$-BanditIV at time T, with probability at least $1-\delta$, is upper-bounded by
\begin{align*}
    R_T &\le B_T \sqrt{2Td \log(\frac{T+d}{d})} + (\frac{2}{\gamma_x}+\|\beta_0\|_2)G_T\sqrt{2Tk \log(\frac{T+k}{k})} + 2\epsilon_0 T L_y
\end{align*}
where 
\begin{align*}
   & B_T\\ &=\sqrt{\gamma_x} ||\beta_0||_2+\sqrt{2\log(\frac{4T}{\delta})+  d \log\left((T \dfrac{k}{d} + 2T^2 k \log (\frac{4d^2T}{\delta}))L_z^2 ||\Gamma_0||_F^2+T L_z  \sqrt{2\log (\frac{4d^2T}{\delta}) } ||\Gamma_0||_1\right) },
\end{align*}
$G_T=\sqrt{\gamma_z} \max_{i\in \{1,\cdots ,d\}}||\Gamma_0^{(i)}||_2+\sqrt{2\log(\frac{2Td}{\delta})+k \log\left(TL_z^2+k\gamma_z \right)}$.
\end{theorem}

\cite{abbasi2011improved} proved an $\tilde{\mathcal{O}}(d\sqrt{T}$) upper bound for the expected regret of OFUL algorithm without considering the endogeneity problem. Taking the endogeneity problem into consideration, we can still keep an upper bound at the same level  with their results by Theorem \ref{thm: total_regret}. 
\begin{remark}
    The upper bound for the expected regret increases with the dimensions of the feature vector and the instrumental variable. Thus, if we have feature vectors of a large dimension, we may need to utilize feature selection techniques to reduce the dimension and hence improve the regret bound in practice.
\end{remark}
\begin{remark}
    Notice that when $\epsilon_t$ increases, the regret bound also increases, which implies that the regret bound of an \textit{$\epsilon$-BanditIV} algorithm with a positive $\epsilon_t$ will always be larger than the regret bound of the \textit{BanditIV} algorithm.
\end{remark}
\begin{remark}
    In order to guarantee an $\tilde{\mathcal{O}}(k\sqrt{T}$) upper bound for the total expected regret of the $\epsilon$-BanditIV algorithm, we need a small enough $\epsilon_t$, such as $\epsilon_t\le  \frac{\sqrt{\log(t)}}{\sqrt{t}}$ or $\epsilon_t\le  \frac{\sqrt{\log\log(t)}}{\sqrt{t}}$ . To see the validity of these examples, note that the total expected regret can be written as upper bounded by $B_T \sqrt{2Td \log(\frac{T+d}{d})} + (\frac{2}{\gamma_x}+\|\beta_0\|_2)G_T\sqrt{2Tk \log(\frac{T+k}{k})} + 2L_y \sum_{t=1}^T \epsilon_t$, where $2L_y \sum_{t=1}^T \epsilon_t$ can be upper bounded by $\tilde{\mathcal{O}}(\sqrt{T})$ if we have $\epsilon_t\le \tilde{\mathcal{O}}(\frac{1}{\sqrt{t}})$. The \textit{BanditIV} algorithm which is a special case with $\epsilon_t=0$ in $\epsilon$-BanditIV algorithm, naturally satisfies this condition.
\end{remark}

% \begin{align}\label{eq: regret_eps_example}
%      \nonumber R_T   &= \sum_{t=1}^T(1-\epsilon_t)\mathbb{E}_{u_t}[\langle x_t^*, \beta_0 \rangle - \langle x_t, \beta_0 \rangle] 1[x_t = \Gamma_0'z_t]        + \sum_{t=1}^T \epsilon_t\mathbb{E}_{u_t}[\langle x_t^*, \beta_0 \rangle - \langle x_t, \beta_0 \rangle] 1[x_t \neq \Gamma_0'z_t] \\ \nonumber &\leq \sum_{t=1}^T\langle \Gamma'_0z_t^*, \beta_0 \rangle - \langle \Gamma'_0z_t, \beta_0 \rangle        + 2L_y \sum_{t=1}^T \epsilon_t \\
%      & \le B_T \sqrt{2Td \log(\frac{T+d}{d})} + (\frac{2}{\gamma_x}+\|\beta_0\|_2)G_T\sqrt{2Tk \log(\frac{T+k}{k})} + 2L_y \sum_{t=1}^T \epsilon_t
% \end{align}

% \begin{align*}
%     \sum_{t=1}^T \frac{1}{\sqrt{t}} \le 2\sqrt{T}-1
% \end{align*}

By setting $\epsilon_t=0$, we can derive the following corollary directly,
\begin{corollary}
    \label{thm: total_regret_eps}
The expected cumulative regret of the BanditIV at time T, with probability at least $1-\delta$, is upper-bounded by
\begin{align*}
    R_T &\le B_T \sqrt{2Td \log(\frac{T+d}{d})} + (\frac{2}{\gamma_x}+\|\beta_0\|_2)G_T\sqrt{2Tk \log(\frac{T+k}{k})}
\end{align*}
% where $B_T=\sqrt{\gamma_x} ||\beta_0||_2+\sqrt{2\log(\frac{2T}{\delta})+ d\log\left(\frac{5TL_x^2}{d} \right)}$,\\ 
% $G_T=\sqrt{\gamma_z} ||\Gamma_0^{(i)}||_2+\sqrt{2\log(\frac{2T}{\delta})+k \log\left(\frac{5TL_y^2}{k} \right)}$.
where $B_T$ and
$G_T$ are the same as those in the Theorem \ref{thm: total_regret}.
\end{corollary}

\begin{lemma}[Optimistic estimation of $\beta_0$]\label{prop:Bt}
With high prob $1-\delta/2$, for all $t \in [T]$,
\begin{align*}
   ||\hat{\beta}_t-\beta_0||_{W_t}
   \le B_t
\end{align*}
where $B_t$ is the same as that in the Theorem \ref{thm: total_regret}.

\end{lemma}

\begin{lemma}[Optimistic estimation of $\Gamma_0$] \label{prop:Gt}
With high prob $1-\delta/2$, for all $t \in [T]$ and all $i \in [d]$,
\begin{align*}
    ||\hat{\Gamma}_t^{(i)}-\Gamma_0^{(i)}||_{U_t}\le G_t
\end{align*}
where $G_t$ is the same as that in the Theorem \ref{thm: total_regret}.
\end{lemma}
 
Due to that we have two-stage estimators in the algorithm, we construct double confidence intervals for the estimators to guarantee the regret bound. In every round, the algorithm chooses estimates for $\Gamma_0$ and $\beta_0$ from the confidence sets as we describe in Section \ref{sec: algorithm}. Lemmas \ref{prop:Gt} and \ref{prop:Bt}show that $\Gamma_0$ and $\beta_0$ with high probability in ellipsoids with center at $\hat{\Gamma}_t$ and $\hat{\beta}_t$ respectively, i.e. in the double confidence sets.

\section{Inference and Confidence Intervals}\label{sec: inference}
Due to the adaptive nature of multi-armed bandits, calculating confidence intervals is not straightforward as the collected data is no longer iid.
In this section, we show asymptotic properties of the \textit{BanditIV} estimator following \cite{chen2021statistical} and \cite{bastani2020online}. We first present the consistency of the online OLS estimator and the \textit{BanditIV} estimator, then we demonstrate the normality of the \textit{BanditIV} estimator. The proofs are presented in Appendix A.

\begin{proposition}(Tail bounds for the online OLS estimators).\label{prop: tail_ols}
    In the online decision making model with the $\epsilon$-greedy policy, if Assumptions \ref{asm: l2norm} and \ref{asm: mineig} are satisfied, and $\epsilon_t$ is non-increasing, then for any $\eta_1, \eta_2>0$, any $i\in \{1,\ldots,d\}$,
    \begin{align*}
          P(||\hat{\delta}_t-\delta_0||_1\le \eta_1)&\ge 1-\exp(-\frac{t\epsilon_t}{8})-k\exp(-\frac{t\epsilon_t\lambda}{32L_z^2})-2k\exp(-\frac{t\epsilon_t^2\lambda^2\eta_1^2}{128k^2\sigma_v^2L_z^2})\\
          &+2k^2\exp(-\frac{t\epsilon_t^2\lambda^2\eta_1^2+4t\epsilon_t\lambda k^2\sigma_v^2}{128k^2\sigma_v^2L_z^2})
    \end{align*}
     \begin{align*}
          P(||\hat{\Gamma}_t^{(i)}-\Gamma_0^{(i)}||_1\le \eta_2)&\ge 1-\exp(-\frac{t\epsilon_t}{8})-k\exp(-\frac{t\epsilon_t\lambda}{32L_z^2})-2k\exp(-\frac{t\epsilon_t^2\lambda^2\eta_2^2}{128k^2\sigma_u^2L_z^2})\\
          &+2k^2\exp(-\frac{t\epsilon_t^2\lambda^2\eta_2^2+4t\epsilon_t\lambda k^2\sigma_u^2}{128k^2\sigma_u^2L_z^2})
    \end{align*}
\end{proposition}

Proposition \ref{prop: tail_ols} states that both the OLS estimator based on the instrumental variable and the first stage OLS estimator are consistent if $t\epsilon_t^2\rightarrow \infty$ as $t\rightarrow \infty$. Based on this result, we derive the tail bound for the \textit{BanditIV} estimator as the following.

\begin{proposition}
\label{lemma: online_beta}
(Tail bound for the BanditIV estimator) In the online decision-making model with $\epsilon$-greedy policy, if the Assumptions \ref{asm: l2norm} and \ref{asm: mineig} are satisfied, and $\epsilon_t$ is non-increasing, then for any $\eta_1, \eta_2> 0$, $\eta_2\neq \frac{\lambda_{min}(\Gamma_0)}{(kd)^{\frac{1}{2}}} $,
    \begin{equation*}
P(||\hat{\beta}_t -\beta_{0}||_1 \le C_{\beta})\ge 1-(p_1+d p_2)
\end{equation*}

where $C_{\beta}=\dfrac{\eta_1+  \eta_2||\beta_0||_1}{\lambda_{min}(\Gamma_0)d^{\frac{-1}{2}}-\eta_2(k)^{\frac{1}{2}}}$, 
\begin{align*}
    p_1= \exp(-\dfrac{t\epsilon_t}{8})+k\exp(-\dfrac{t\epsilon_t\lambda}{32L_z^2})+2k\exp(-\dfrac{t\epsilon_t^2\lambda^2\eta_1^2}{128k^2\sigma_v^2L_z^2})-2k^2\exp(-\dfrac{t\epsilon_t^2\lambda^2\eta_1^2+4t\epsilon_t\lambda k^2\sigma_v^2}{128k^2\sigma_v^2L_z^2}),
\end{align*}
\begin{align*}
    p_2=\exp(-\dfrac{t\epsilon_t}{8})+k\exp(-\dfrac{t\epsilon_t\lambda}{32L_z^2})+2k\exp(-\dfrac{t\epsilon_t^2\lambda^2\eta_1^2}{128k^2\sigma_u^2L_z^2})
-2k^2\exp(-\dfrac{t\epsilon_t^2\lambda^2\eta_1^2+4t\epsilon_t\lambda k^2\sigma_u^2}{128k^2\sigma_u^2L_z^2})
\end{align*}

\end{proposition}
\begin{remark}
Following the Proposition \ref{prop: tail_ols}, if $t\epsilon_t^2 \rightarrow \infty$ as $t\rightarrow \infty$, then the probability of $||\hat{\beta}_t -\beta_{0}||\le C_{\beta}$ goes to 1 for any $\eta_1, \eta_2> 0$, $\eta_2\neq \frac{\lambda_{min}(\Gamma_0)}{(kd)^{\frac{1}{2}}} $. Combining the Theorem \ref{thm: total_regret} and Proposition \ref{lemma: online_beta}, in order to guarantee both the regret bound and the tail bound, we need to carefully choose the value for $\epsilon_t$. Some reasonable examples can be $\epsilon_t=  \frac{\sqrt{\log(t)}}{\sqrt{t}}$ or $\epsilon_t=  \frac{\sqrt{\log\log(t)}}{\sqrt{t}}$.
\end{remark}
\begin{remark}
    The probability of $||\hat{\beta}_t -\beta_{0}||\le C_{\beta}$ decreases with the dimension of instrumental variables, the variance of error terms in both stages of regression model, and $L_z$. Moreover, the probability of $||\hat{\beta}_t -\beta_{0}||\le C_{\beta}$ increases with the lower bound of the minimum eigenvalue of $\mathbb{E}[z_{t,a}z'_{t,a}]$.
\end{remark}

Following from the Propositions \ref{prop: tail_ols} and \ref{lemma: online_beta}, we obtain the consistency of the OLS estimators $\hat{\delta}_t$ and $\hat{\Gamma}_t$ and the BanditIV estimator $\hat{\beta}_t$ easily.
\begin{corollary}\label{cr: consistency}
(Consistency of the online OLS estimators). If Assumptions \ref{asm: l2norm} and \ref{asm: mineig} are satisfied, $\epsilon_t$ is non-increasing and $t\epsilon_t^2 \rightarrow \infty$ as $t\rightarrow \infty$, then the online OLS estimators $\hat{\Gamma}_t$ is a consistent estimator for $\Gamma_0$ and $\hat{\delta}_t$ is a consistent estimator for $\delta_0$.
\end{corollary}

\begin{corollary}
(Consistency of the online BanditIV estimator). If Assumptions \ref{asm: l2norm} and \ref{asm: mineig} are satisfied, $\epsilon_t$ is non-increasing and $t\epsilon_t^2 \rightarrow \infty$ as $t\rightarrow \infty$, then the online BanditIV estimator $\hat{\beta}_t$ is a consistent estimator for $\beta_0$.
\end{corollary}

\begin{theorem}\label{thm: normality}
(Aymptotic normality of the online OLS estimator)
 If Assumptions \ref{asm: l2norm} and \ref{asm: mineig} are satisfied, then
 \begin{align*}
     \sqrt{t}(\hat{\delta}_t-\delta_0) \xrightarrow{d} \mathcal{N}_k(0,\mathbb{E}[v_t^2](\int zz'd\mathcal{P}_z)^{-1})
 \end{align*}
\end{theorem}

 Based on above results, we can derive the normality of the BanditIV estimator by applying Slutsky Theorem. We also provide a consistent estimator for the variance of the normal distribution. The proof for the  consistency of the variance estimator can be found in Appendix A.
\begin{theorem}(Asymptotic normality of the online BanditIV estimator)\label{thm: asymnorm_banditiv}
If Assumptions \ref{asm: l2norm} and \ref{asm: mineig} are satisfied, $\epsilon_t$ is non-increasing and $t\epsilon_t^2 \rightarrow \infty$ as $t\rightarrow \infty$. Then
\begin{align*}
    \sqrt{t}(\hat{\beta}_t - \beta_0) \xrightarrow{d} \mathcal{N}_d(0,S)
\end{align*}
where $S=\mathbb{E}[v^2](\Gamma'_0 \int zz'd\mathcal{P}_z \Gamma_0)^{-1}\Gamma'_0 \int zz'd\mathcal{P}_z  \Gamma_0(\Gamma'_0 \int zz'd\mathcal{P}_z \Gamma_0)^{-1}$

A consistent estimator for $S$ is given by
\begin{align*}
   \sum_{s=1}^t \hat{v}_s^2(\hat{\Gamma}'_t \sum_{s=1}^t z_sz'_s \hat{\Gamma}_t)^{-1}\hat{\Gamma}'_t(\sum_{s=1}^t z_sz'_s)\hat{\Gamma}_t(\hat{\Gamma}'_t \sum_{s=1}^t z_sz'_s \hat{\Gamma}_t)^{-1}
\end{align*}
where $\hat{v}_s=y_s - (\hat{\Gamma}_s\hat{\beta}_s)'z_s$.
\end{theorem}
% \begin{proof}
%      We know that 
%      \begin{align*}
%          \hat{\beta}_t=(\hat{\Gamma}'_t\hat{\Gamma}_t)^{-1}\hat{\Gamma}'_t\hat{\delta}_t
%      \end{align*}
     
%      Combining the Theorem \ref{thm: normality} and Corollary \ref{cr: consistency} by applying Slutsky Theorem, we complete the proof.
     
% \end{proof}

Theorem \ref{thm: asymnorm_banditiv} provides a theoretical guarantee that the \textit{BanditIV} estimator asymptotically follows a normal distribution with mean zero and variance $S$. We can see that this variance depends on the expectation of the error term $v_t$, which includes the errors in both first stage and the second stage. The variance also depends on the distribution of the instrumental variable, $\mathcal{P}_z$. Notice that, although we need the assumption about $\epsilon_t$ to guarantee the consistency of the estimator, the asymptotic variance of the \textit{BanditIV} estimator does not depend on $\epsilon_t$.

 To compute confidence intervals of our estimate of $\beta_0$ post running our bandits, we provide two ways as the following. As the first way, we use directly Theorem \ref{thm: asymnorm_banditiv}. The data set is constructed as illustrated in Section \ref{sec: simulation}. We set all elements in $\Gamma_0$ and $\beta_0$ as $1$. We run 1000 trials in total. In each trial, we take the final estimation $\hat{\beta}$ from a pre-run $\epsilon$-BanditIV with 1000 rounds as the mean of the confidence interval and calculate the estimated standard deviation as the theorem provides to construct one confidence interval. Among the 1000 trials, for $95\%$ confidence intervals, we get $94.4\%$ coverage when we set $k=2$, $d=1$ and $94.6\%$ coverage when we set $k=1$, $d=1$, which are very close to the ideal.
 
 As the second way, we use a re-randomization test similar to that of \cite{bojinov2020design} and  \cite{farias2022synthetically}. For the estimate $\hat{\beta}$, we test the sharp null hypothesis that the $\beta_0=\tau$ for all $t$. The sharp null hypothesis implies that the new outcome is $y_t+\tau'(x^{H}_t-x_t)$ where $x^{H}_t$ is a newly pulled arm and $x_t$ is a pulled arm by our bandit algorithm.

We conduct exact tests by using the known assignment mechanism to simulate new assignment
paths. Algorithm \ref{sharp null Test} provides the details of how to implement it. In particular, we propose $\tau$ by a downward search method based on the estimate from the bandit algorithm. Under the sharp
null hypothesis of $\beta_0=\tau$, a new arm assignment path
$1:T$ leads to a sequence of observed outcomes $y_t+\tau'(x^{H}_t-x_t)$ for $t\in \{1,\cdots ,T\}$. To obtain a confidence
interval, we invert a sequence of exact hypothesis tests to identify the region outside where the null hypothesis is violated at the prespecified significance level \citep{imbens2015causal, bojinov2020design}. 

\begin{algorithm}[H]
\caption{{\bf Sharp Null Hypothesis Test} \label{sharp null Test}}
\begin{algorithmic}
\Require Fix $N_H$, total number of samples drawn; Given $\hat{\Gamma}$, $X$, and $Y$ from the bandit algorithm; Given a prespecified significance level $\alpha$
\For{$i$ in $1:N_H$}
\State Sample a new assignment path, $x^H_t$, for $t\in \{1,\cdots ,T\}$ according to the assignment mechanism under the null hypothesis $\beta_0=\tau$
\State Change the sequence of original outcomes $y_t$ to $y_t+\langle \tau,(x^H_t-x_t)\rangle$ for $t\in \{1,\cdots ,T\}$
\State Compute $\hat{\tau}^{[i]}$ as the original algorithm does, i.e.
\begin{equation*}
        \hat{X}^H = Z^H\hat{\Gamma}
\end{equation*}
\begin{equation*}
    \hat{\tau}^{[i]}=\gamma I + ((\hat{X}^H)'\hat{X}^H)^{-1}(\hat{X}^H)'{Y}^H
\end{equation*}
\EndFor
\State Compute $\hat{p}=N_H^{-1}\sum_{i=1}^{N_H}\bf{1}\{|\hat{\tau}^{[i]}|>|\hat{\beta}|\}$
\State Reject the null hypothesis if $\hat{p}<\alpha$
\end{algorithmic}
\end{algorithm}

Consistent with the setting illustrated in the first way, all of these experiments are run on synthetic data constructed as in Section \ref{sec: simulation} and we consider two cases including $k=2$, $d=1$, and $k=1$, $d=1$. For each case, we run 50 trials. In each trial, we propose 20 hypotheses for $\beta_0$ based on the estimated coefficient of our main interest, $\hat{\beta}$ from a pre-run $\epsilon$-BanditIV and construct confidence intervals for all dimensions of $\beta_0$. For each hypothesis of $\beta_0$, we run the sharp null hypothesis test where we set $N_H$ as 200.  We choose the significance level to be $\alpha=0.05$. Under $\epsilon$-greedy BanditIV, the coverage of the test is 96\% when $k=2$, $d=1$ and 94\% when $k=1$, $d=1$. We can see that, given the significance level as $0.05$, the coverage of the test is close to ideal. Thus, the re-randomization tests and corresponding confidence intervals reported here are adequate for inference of the main treatment effect.

\section{Numerical Experiments}\label{sec: simulation}
In this section, we construct synthetic data to further validate our algorithm. Referring the simulation set up in \cite{bakhitov2021causal}, we consider the model as follows, for $t=1,\cdots, T$,
\begin{align*}
    Y_t=X_t\beta_0+e_t\rho + \varepsilon_t\\
    X_t=Z_t\Gamma_0+e_t + u_t
\end{align*}

Suppose that all elements of the instrument $Z_t \in \mathbb{R}^{k \times 1}$ are uniformly distributed on the support $[-3,3]$. The error term $e_t\in \mathbb{R}^{n_t \times d}$ is the confounder, where $n_t$ is the number of arms at time $t$ and all elements of $e_t$ follows $\mathbb{N}(0,1)$. The parameter $\rho \in \mathbb{R}^{d \times 1}$ measures the degree of endogeneity. A lower $\rho$ implies a less serious endogeneity issue. As an extreme example, when $\rho=0$, endogeneity disappears, which we can see from Equation (\ref{eq:endogeneity}). The additional noise terms $u_t\in \mathbb{R}^{n_t \times d}$, and $\varepsilon_t\in \mathbb{R}^{n_t \times 1}$ are i.i.d. normally distributed. All elements of $u_t$ and $\varepsilon_t$ follow $\mathbb{N}(0,0.01)$ and $\mathbb{N}(0,1)$ respectively. WLOG, we set $n_t=50$, for $t=1,\cdots, T$ in this simulation.
\begin{align}\label{eq:endogeneity}
    \mathbb{E}[(e_t\rho + \varepsilon_t)'X_t]=\mathbb{E}[\rho'e'_tX_t]= \mathbb{E}[\rho'e'_t \mathbb{E}[X_t|e_t]]=\rho'\mathbb{E}[e'_te_t]
\end{align}

\begin{figure}
   \begin{subfigure}{0.5\textwidth}
        \centering
        \includegraphics[scale=0.4]{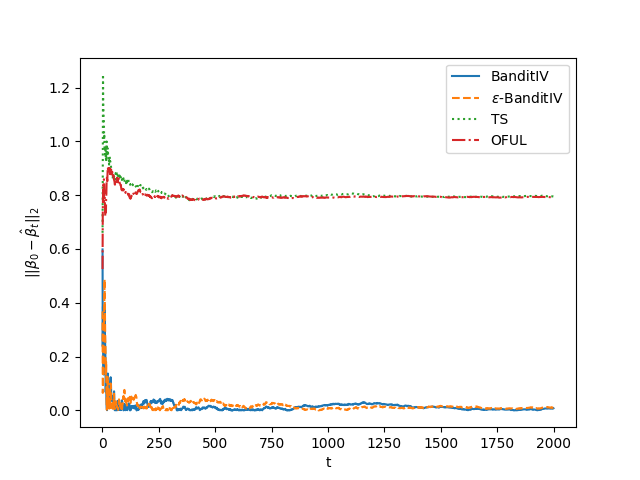}
        \caption{Estimation bias}
        \label{fig: bias_r2}
    \end{subfigure}
    \hfill
   \begin{subfigure}{0.5\textwidth}
        \centering
        \includegraphics[scale=0.4]{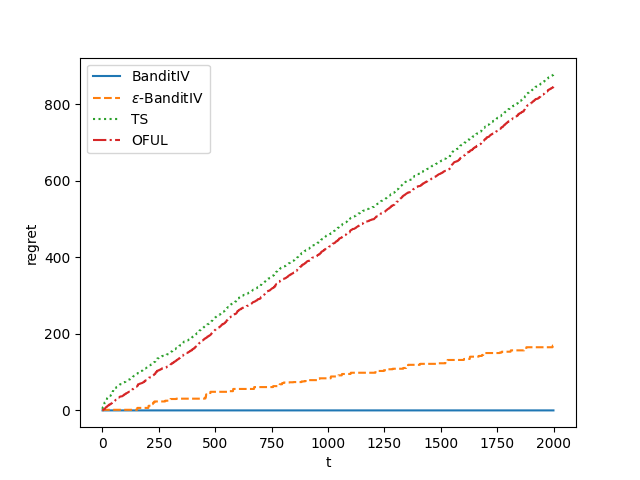}
        \caption{Regret variations}
        \label{fig: regret_r2}
    \end{subfigure}
    \vfill
     \begin{subfigure}{0.5\textwidth}
        \centering
        \includegraphics[scale=0.4]{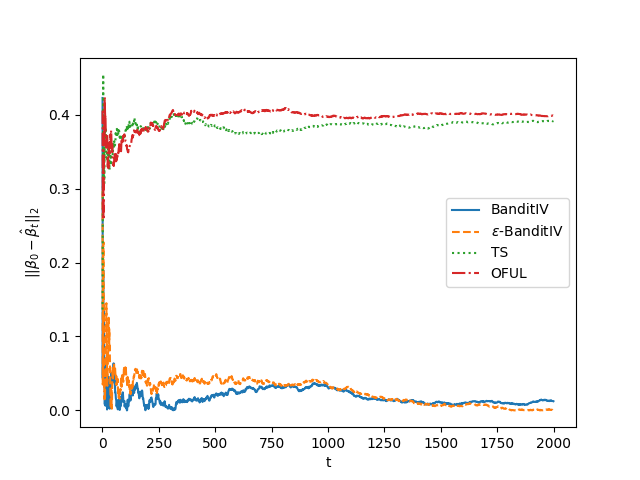}
        \caption{Estimation bias}
        \label{fig: bias_r1}
    \end{subfigure}
    \hfill
   \begin{subfigure}{0.5\textwidth}
        \centering
        \includegraphics[scale=0.4]{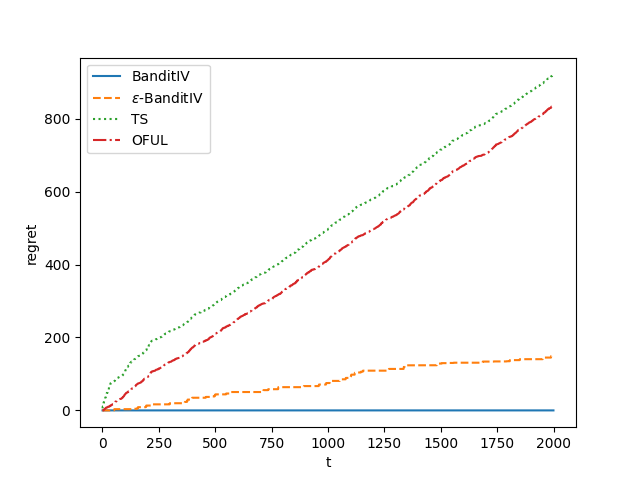}
        \caption{Regret variations}
        \label{fig: regret_r1}
    \end{subfigure}
    \vfill
     \begin{subfigure}{0.5\textwidth}
        \centering
        \includegraphics[scale=0.4]{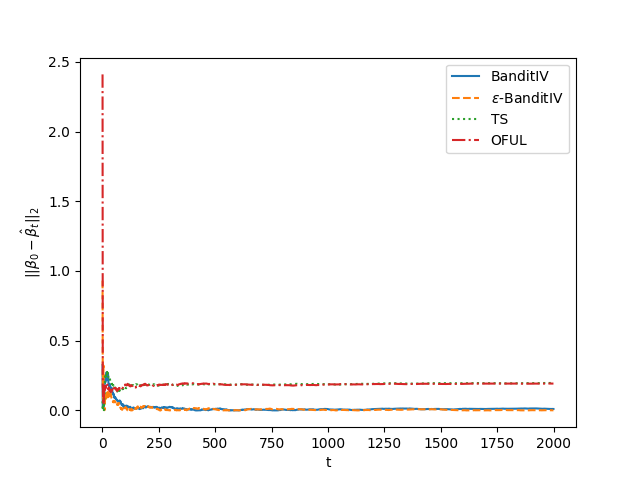}
        \caption{Estimation bias}
        \label{fig: bias_r0.5}
    \end{subfigure}
    \hfill
   \begin{subfigure}{0.5\textwidth}
        \centering
        \includegraphics[scale=0.4]{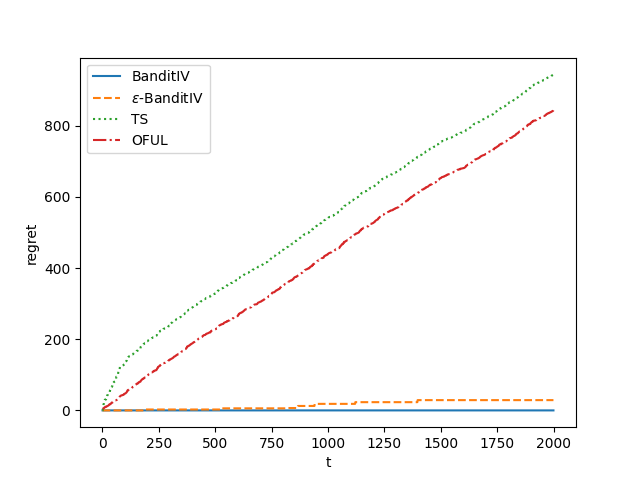}
        \caption{Regret variations}
        \label{fig: regret_r0}
    \end{subfigure}
    \caption{Simulation results on synthetic data with endogeneity when $k=1, d=1$}
    \caption*{\small The x-axis shows the number of time steps and the y-axis shows the performance indicator which is either estimation bias or the regret. Subfigures (a)-(b) present results when $\rho=2$, (c)-(d) present results when when $\rho=1$, (e)-(f) present results when when $\rho=0.5$}
    \label{fig:synthetic}
\end{figure}

\begin{figure}
     \begin{subfigure}{0.5\textwidth}
        \centering
        \includegraphics[scale=0.4]{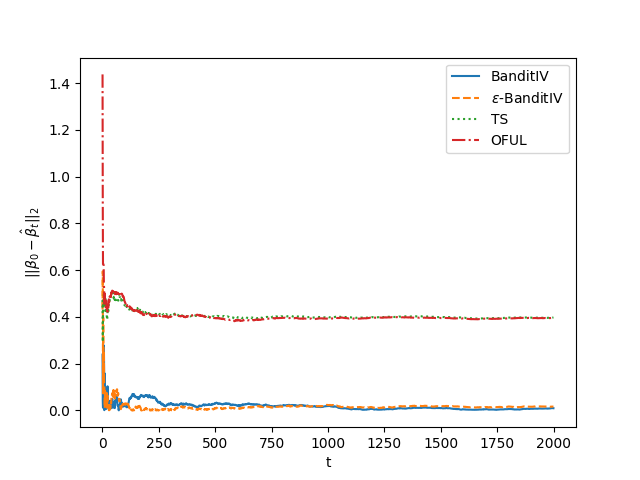}
        \caption{Estimation bias}
        \label{fig: 2bias_r2}
    \end{subfigure}
    \hfill
   \begin{subfigure}{0.5\textwidth}
        \centering
        \includegraphics[scale=0.4]{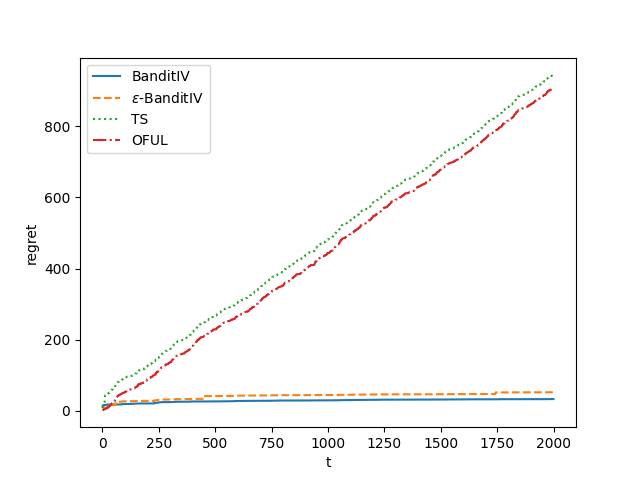}
        \caption{Regret variations}
        \label{fig: 2regret_r2}
    \end{subfigure}
    \vfill
     \begin{subfigure}{0.5\textwidth}
        \centering
        \includegraphics[scale=0.4]{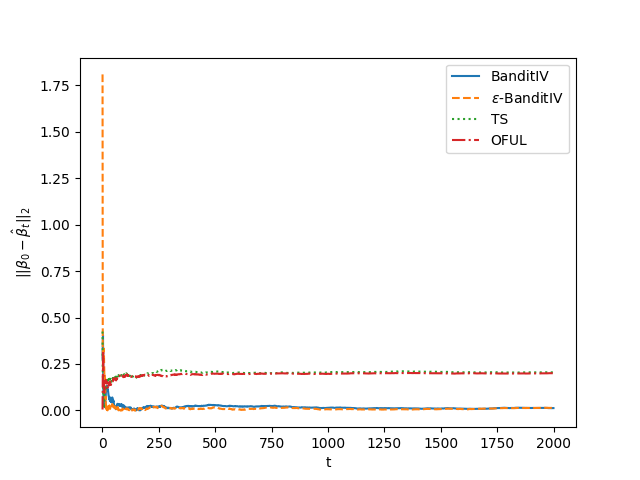}
        \caption{Estimation bias}
        \label{fig: 2bias_r1}
    \end{subfigure}
    \hfill
   \begin{subfigure}{0.5\textwidth}
        \centering
        \includegraphics[scale=0.4]{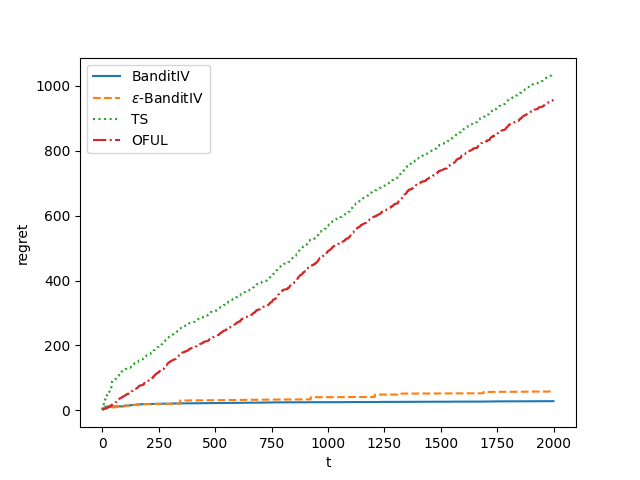}
        \caption{Regret variations}
        \label{fig: 2regret_r1}
    \end{subfigure}
    \vfill
    \begin{subfigure}{0.5\textwidth}
        \centering
        \includegraphics[scale=0.4]{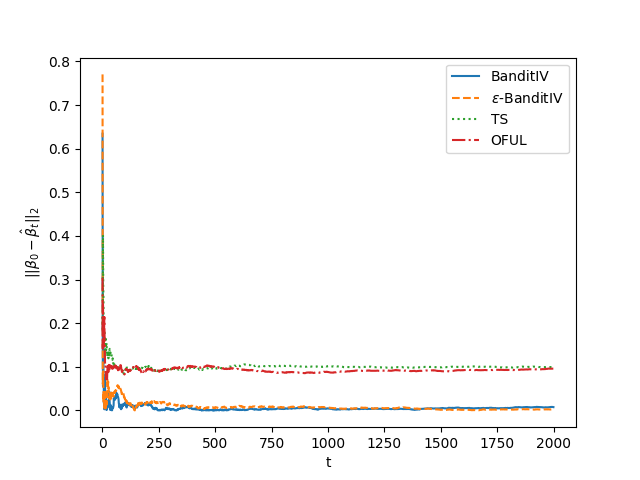}
        \caption{Estimation bias}
        \label{fig: 2bias_r0.5}
    \end{subfigure}
    \hfill
   \begin{subfigure}{0.5\textwidth}
        \centering
        \includegraphics[scale=0.4]{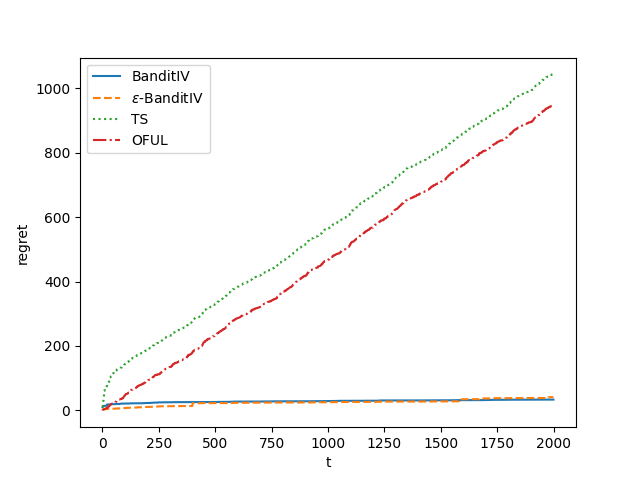}
        \caption{Regret variations}
        \label{fig: 2regret_r0}
    \end{subfigure}
    \caption{Simulation results on synthetic data with endogeneity when $k=2, d=1$}
     \caption*{\small The x-axis shows the number of time steps and the y-axis shows the performance indicator which is either estimation bias or the regret. Subfigures (a)-(b) present results when $\rho=2$, (c)-(d) present results when when $\rho=1$, (e)-(f) present results when when $\rho=0.5$}
    \label{fig:synthetic2}
\end{figure}

We have two objectives in the simulation: (i) to achieve the maximum reward through selecting optimal arms (ii) to obtain an accurate estimation of the causal relation parameter $\beta_0$. We compare performance of our algorithm regarding the two objectives, with other existing algorithms including TS and OFUL under endogeneity. We run $T=2000$ time steps for each algorithm and observe the regret and estimation bias along the time. We use the true cumulative regret which excludes random error terms to measure the regret and $||\beta_0-\hat{\beta}_t||_2$ to measure the estimation bias. Figures \ref{fig:synthetic} and \ref{fig:synthetic2} show the results where we consider various endogeneity degrees and dimensions of the instrumental variable. In Figure \ref{fig:synthetic}, we set the dimension of the instrumental variable as $k=1$ which is equal to the dimension of endogenous variable $d=1$. When $k=d$, we have the same number of instrumental variables as that of endogenous variables and $\beta_0$ will be just identified. In Figure \ref{fig:synthetic2}, we set the dimension of the instrumental variable as $k=2$ which is larger then the dimension of endogenous variable $d=1$. When $k>d$, we have more instrumental variables than endogenous variables, which can cause overidentification.  Across these two cases regarding the dimensions, we find  quite robust results that our proposed algorithms outperform TS and OFUL both on regret and inference. This outperformance is more significant under higher endogeneity. In both Figures \ref{fig:synthetic} and \ref{fig:synthetic2}, the first, second, third row present results when $\rho=2, 1, 0.5$ respectively. We can see that BanditIV and $\epsilon$-{BanditIV} can achieve lower bias in inference than TS and OFUL and this difference become larger when the endogeneity degree increases. Also, BanditIV reaches lower expected regret than $\epsilon$-BanditIV, but $\epsilon$-BanditIV can obtain a less biased estimation.

\section{Real-Time Bidding Auctions}\label{sec: rtb}
RTB is a dominant auction mechanism for selling ad exposures in display markets. A consumer entering a publisher's website generates an ad impression opportunity for advertisers. An auction takes place for a single impression opportunity. In RTB, an advertiser makes buying decision in the auction immediately after the consumer arrives on the publisher's website. Buying decisions of impressions involve programmatic automation. The advertiser's ad is served to the consumer only if the advertiser wins the auction of the impression. 

An advertiser is able to make buying decisions based on impression-specific information, like the width, height and visibility of an ad slot, and the user's profile. Therefore, we can model the advertiser's buying problem as a linear bandit problem. The objective of this problem is twofold: first, to measure the marginal value of the ad exposure; second, to achieve the largest profit through optimizing the bidding price. Notice that the advertiser has to bid in an auction to win the impression opportunity. Either overbidding or underbidding may bring significant costs to the advertiser. The advertiser can obtain a higher winning probability with a very high bid but may not cover the paying price with the profit of ad exposure. On the contrary, the advertiser will be less likely to win with a low bid.

Now, we illustrate the  model (see Equation (\ref{eq: rtb_revenue1})) and explain the potential endogeneity issue. We model the advertiser's revenue ($\Tilde{y}_t$) from winning the auction as some base revenue ($\beta_0$) plus the revenue generated from the ad exposure ($\beta_1$). The advertiser profit ($y_t$) is given by the advertiser's revenue minus the paying price in case the advertiser wins the auction ($\mathbb{I}\{B_{CP,t}\le b_t\}$).  For our setting, we assume a second-price auction. The base revenue is a part of the revenue the advertiser will always get no matter whether the advertiser wins the auction or not. We use $\beta_0$ to represent the expected base revenue. The marginal profit of the ad exposure, $\beta_1$, is our main coefficient of interest and we also call it as the  treatment effect i.e., the effect of ad exposures on revenue. We use an indicator function, $\mathbb{I}\{B_{CP,t} \le b_t\}$, to present whether the advertiser wins the auction or not, where $B_{CP,t}$ is the highest bidding price among the advertiser's competitors at time $t$ and $b_t$ is the advertiser's bidding price at time $t$. Notice that $B_{CP,t}$ is also the advertiser's paying price in the second-price auction. We use $\eta_t$ to denote the unobserved demand shocks at time $t$.
% Following \cite{waisman2019online} , we write the function of advertiser's revenue as Equation (\ref{eq: rtb_revenue0}), where $\mathbb{E}[\gamma(0)]$ is the expected basic revenue, $\mathbb{E}[\gamma(1)-\gamma(0)]$ is the expected revenue generated from the ad exposure, and $\eta_t$ is a normal random error term.

% \begin{equation}\label{eq: rtb_revenue0}
%     y_t = \mathbb{E}[\gamma(0)] + (\mathbb{E}[\gamma(1)-\gamma(0)])\times \mathbb{I}\{B_{CP,t} \le b_t\} + \eta_t
% \end{equation}

% \cite{waisman2019online} proposed bandit algorithm to recover $\mathbb{E}[\gamma(1)-\gamma(0)]$ based on the assumption of exogenous covariates including $\mathbb{I}\{B_{CP,t} \le b_t\}$. However, we can see this assumption will be violated in reality.

% \begin{equation}\label{eq: rtb_profit}
% y_t=\beta_0 + (\beta_1) \mathbb{I}\{B_{CP,t}\le b_t\} + \eta_t
% \end{equation}

\begin{equation}\label{eq: rtb_revenue1}
\Tilde{y}_t=\beta_0 + \beta_1 \mathbb{I}\{B_{CP,t}\le b_t\} + \eta_t
\end{equation}
% {\color{red} What is y? "I couldn't find what are x, z and y in the RTB example"}
Note that the above specification suffers from the endogeneity problem, i.e. the covariate $\mathbb{I}\{B_{CP,t} \le b_t\}$ can be correlated with the unobserved error term $\eta_t$. The advertiser cannot randomize the ad exposure because the advertiser does not know others' bids, and the result of the auction cannot be controlled by the advertiser; Thus, the unobserved demand shocks could be  correlated with $B_{CP,t}$ and create a potential endogeneity issue. The unobservables here could be some common shock in the market, which affects the bidding prices in the market but are not observable in the data in the error term $\eta_t$. Extant literature (\cite{johnson2017ghost}) has characterized this inherent endogeneity in RTB settings and discusses how unobserved consumer taste shocks affect both consumer demand and advertisers' bidding behavior. 

% In this paper, we relax the assumption of exogenous covariates to accommodate more general settings which include endogenous cases.
To make the analysis clearer, we reformulate the error term in the model as Equation (\ref{eq: rtb_revenue_err}), where we rewrite $\eta_t$ as the sum of two errors, $\delta_t$ and $\varepsilon_t$. We assume $\mathbb{I}\{B_{CP,t} \le b_t\}$ is correlated with the first error term, $\delta_t$, while uncorrelated with the second error term, $\varepsilon_t$. The parameter $\rho$ indicates the degree of endogeneity.  Larger $\rho$ implies higher endogeneity.
\begin{align}\label{eq: rtb_revenue_err}
    \eta_t=\rho\delta_t +\varepsilon_t
\end{align}

To correct the endogeneity-generated bias of estimation for $\beta_1$, we can utilize instrumental variable methods. A valid instrumental variable $\nu_t$ for $\mathbb{I}\{B_{CP,t} \le b_t\}$ should satisfy the following conditions: (i) $\mathbb{E}[\nu_t\eta_t]=0$, (ii) $\nu_t$ affects $\mathbb{I}\{B_{CP,t} \le b_t\}$, (iii) $\nu_t$ has no direct effect on $y_t$. 
% A few examples of potential exogenous data that can be used as instrument variables in RTB auctions could be -- exchange rates (\cite{clements2005indirect}) and political ads (\cite{sinkinson2019ask}) can be used as instrument variables in RTB auctions. 
% We can use exchange rates as instruments in the RTB problem because many competing advertising companies in RTB auctions could be multinationals with headquarters not the same as the focal company. Thus arguably the exchange rate could affect competing companies advertising budgets and hence bids. Further consumer behavior in the focal country might not be affected as much by the foreign exchange rates (This is particularly true if the focal country is the united states).
One example of potential exogenous data that can be used as instrument variables in RTB auctions could be -- the intensity of political ads (\cite{sinkinson2019ask}) at a given time. Political ads have a displacement effect on commercial ads (\cite{sinkinson2019ask}). Yet, political ads don't have a direct effect on the revenue generated from commercial ads. Therefore, the variation of political ads can represent the exogenous variation in the advertiser winning an auction and serve as a potential instrument in the RTB problem. For our exercise, we artificially simulate such an instrument and assume a linear relationship between  $\mathbb{I}\{B_{CP,t} \le b_t\}$ and the instrument ($\nu_t$).
% Similarly, political ads have a displacement effect on commercial ads (\cite{sinkinson2019ask}). Yet, political ads don't have a direct effect on the revenue generated from commercial ads. Therefore, the variation of political ads can represent the exogenous variation in commercial ads and we can use political ads as an instrument in the RTB problem.
{
% \color{red}

\begin{comment}
Hence, we can use the instrumental variable to extract the exogenous part of endogenous covariate and conduct two stage estimation by our $\epsilon$-BanditIV algorithm. 
% {\color{red} Write this in english (to do)}
In the first stage, we regress $\mathbb{I}\{B_{CP,t} \le b_t\}$ on the impression's characteristics, the bidding price $b_t$, and the instrumental variable, $\nu_t$ to obtain an estimation of $\Gamma$. Based on the estimation $\hat{\Gamma}$, we estimate  $\hat{\mathbb{I}}\{B_{CP,t} \le b_t\}$ as an exogenous part of the original $\mathbb{I}\{B_{CP,t} \le b_t\}$, through $\hat{\Gamma}'z_t$. In the second stage, we regress the revenue on the $\hat{\mathbb{I}}\{B_{CP,t} \le b_t\}$ to get the estimation of $\beta_1$ as well as $\beta_0$ (we denote $\beta=[\beta_0 \quad \beta_1]$). After the estimation, we identify the best bid which maximizes the estimated profit jointly with a pair of most optimistic $\Gamma$ and $\beta$ in the double confidence sets. With probability $\epsilon_t$, we choose a uniformly random bid from the bid set; with probability $1-\epsilon_t$, we choose the best bid.
\end{comment}

% , i.e.
 
% Following the notations in \cite{waisman2019online}, we denote the revenue the advertiser gains when their ad is shown to consumers as $\gamma(1)$ and when their ad is not shown as $\gamma(0)$. The treatment effect of ad exposure is thus $\gamma(1)-\gamma(0)$, which is the coefficient of interest. We can model the revenue gained by an advertiser as the following,

To evaluate the performance of our proposed $\epsilon$-\textit{BanditIV} algorithm on this RTB problem, we use real data from iPinYou \citep{liao2014ipinyou}. The iPinYou RTB data set is publicly available and includes logs of ad biddings, impressions, clicks, and final conversions. We use the impression data for one product category, \emph{Telecom}. Impression slot characteristics, bidding price, and paying price are main variables we use in this section. 

 An advertiser has 2000 time periods to run in total. In each time period, the advertiser is shown an impression opportunity. The advertiser observes the characters of the impression slot. The advertiser has to estimate competitors' bidding prices and decide a bidding price to participate in the auction of the impression. We assume the advertiser considers a set of $R$ arms $r=1,\cdots, R$, each of which associated with a bidding price, $b_{r}$. WLOG, we let $b_{1}<b_{2}<\ldots<b_{R}$ for all $x$. We use the same set of arms for all contexts for simplicity; this setup can trivially be adjusted to accommodate context-specific arms and context-specific number of arms. After the advertiser bids with the chosen arm, $b_r$, the advertiser can observe whether wins or not, paying price for the auction and the realized profit.
 
% Assuming the advertiser is in a second price auction, her observed profit $\pi_t$ can be written as Equation (\ref{eq: rtb_profit}).

% remove eq:

% \begin{equation}\label{eq: rtb_profit}
%     \pi_t=\beta_0 + (\beta_1-B_{CP,t}) \mathbb{I}\{B_{CP,t}<b_t\} + \rho\delta_t +\varepsilon_t
% \end{equation}

Applying the $\epsilon$-BanditIV algorithm to the RTB problem, we estimate the treatment effect of ad exposure based on past records of impression characteristics, bidding prices, instrumental variables, whether win or not, and rewards. We choose the bidding price which maximizes the estimated profit jointly with optimistic estimates of coefficients within the double confidence sets of two-stage coefficients.
% , as Equation (\ref{eq: rtb_arm}).

Figures \ref{fig:rtb1__}, \ref{fig:rtb1} and \ref{fig:rtb1_} show the results of regret variation and estimation bias for four algorithms including BanditIV, $\epsilon$-{BanditIV}, TS and OFUL on the RTB data. We compare algorithm performance under endogeneity case when the endogeneity degree changes. We find that BanditIV and $\epsilon$-BanditIV significantly outperform other two classic algorithms both on regret and inference, especially under higher endogeneity cases. In Figure \ref{fig:rtb_bias02__} and \ref{fig:rtb_bias02}, at round $2000$, TS and OFUL still have larger than 3 times the norm difference of the $\epsilon$-BanditIV or BanditIV; in Figure \ref{fig:rtb_bias02_}, at round $2000$, TS and OFUL have around than 2.7 times the norm difference of the $\epsilon$-BanditIV or BanditIV. Regarding the regret variation, there are slight differences between the performances of TS and OFUL, or between BanditIV and $\epsilon$-{BanditIV}. But due to the large magnitude of numbers, we can barely see the differences. To conclude, our proposed algorithms consistently outperform state-of-art algorithms like TS and OFUL, in the aspects of regret and inference.

\begin{figure}[H]
\begin{subfigure}{0.3\textwidth}
        \centering
        \includegraphics[scale=0.3]{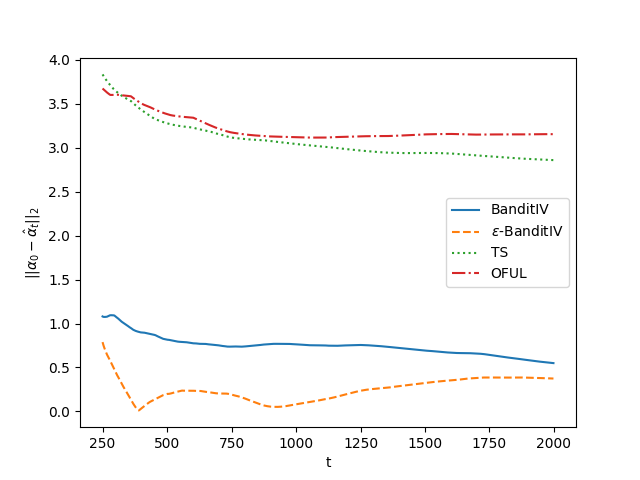}
        \caption{Estimation bias} 
        \label{fig:rtb_bias02__}
    \end{subfigure}
    \hfill
    \begin{subfigure}{0.3\textwidth}
        \centering
        \includegraphics[scale=0.3]{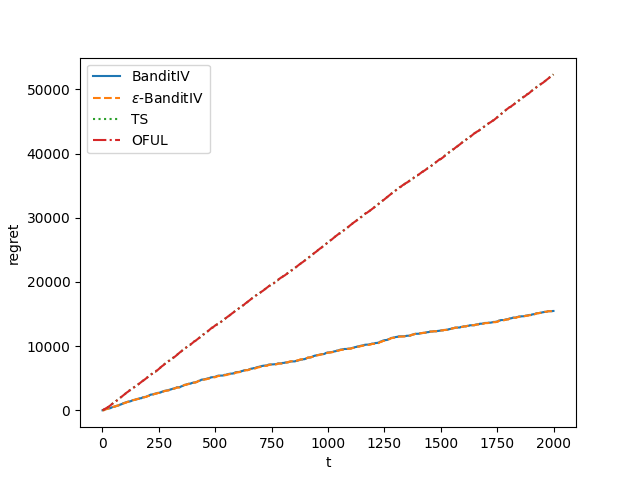}
        \caption{Regret variations} 
        \label{fig:rtb021__}
    \end{subfigure}
    \hfill
     \begin{subfigure}{0.3\textwidth}
        \centering
        \includegraphics[scale=0.3]{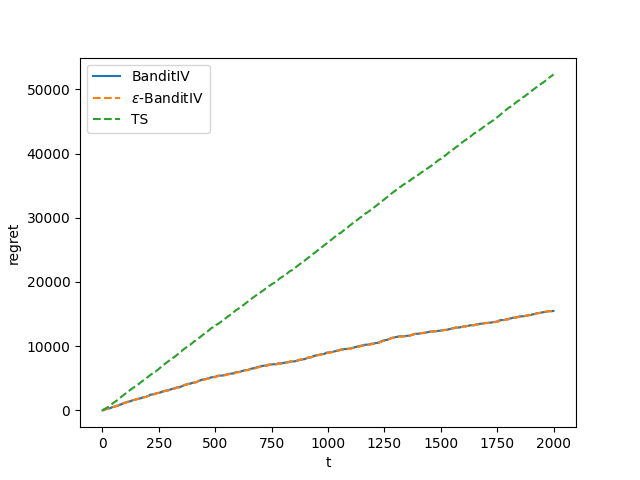}
        \caption{Regret variations} 
        \label{fig:rtb022__}
    \end{subfigure}
    
    \caption{Simulation results on RTB data with endogeneity: $\rho=2$}
    \label{fig:rtb1__}
\end{figure}

\begin{figure}[H]
    \begin{subfigure}{0.3\textwidth}
        \centering
        \includegraphics[scale=0.3]{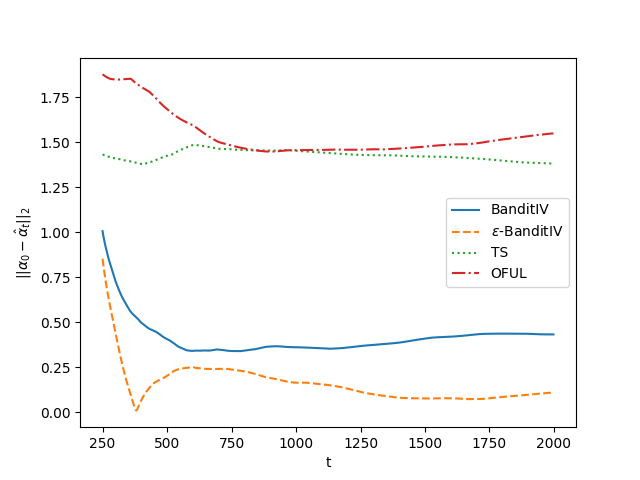}
        \caption{Estimation bias} %  when $\epsilon=0.0005$
        \label{fig:rtb_bias02}
    \end{subfigure}
    \hfill
    \begin{subfigure}{0.3\textwidth}
        \centering
        \includegraphics[scale=0.3]{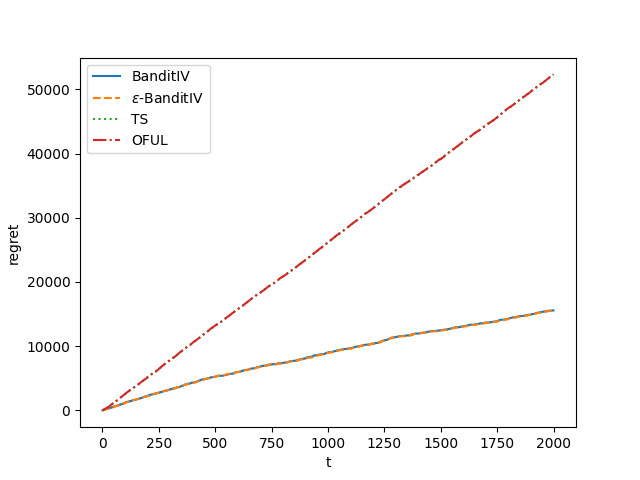}
        \caption{Regret variations} 
        \label{fig:rtb021}
    \end{subfigure}
    \hfill
     \begin{subfigure}{0.3\textwidth}
        \centering
        \includegraphics[scale=0.3]{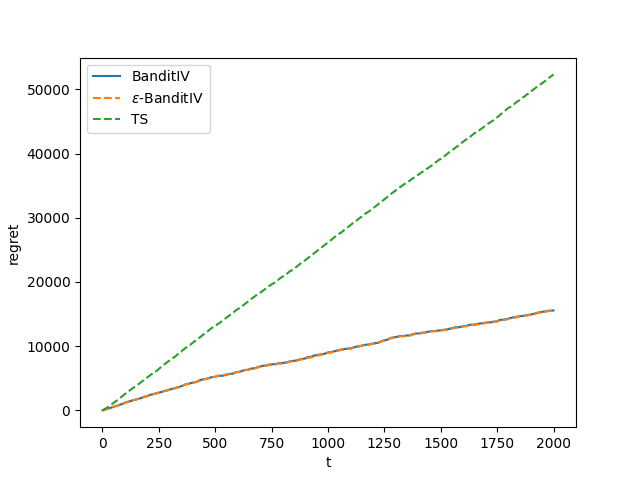}
        \caption{Regret variations} 
        \label{fig:rtb022}
    \end{subfigure}
    \caption{Simulation results on RTB data with endogeneity: $\rho=1$}
    \label{fig:rtb1}
\end{figure}

\begin{figure}[H]
    \begin{subfigure}{0.3\textwidth}
        \centering
        \includegraphics[scale=0.3]{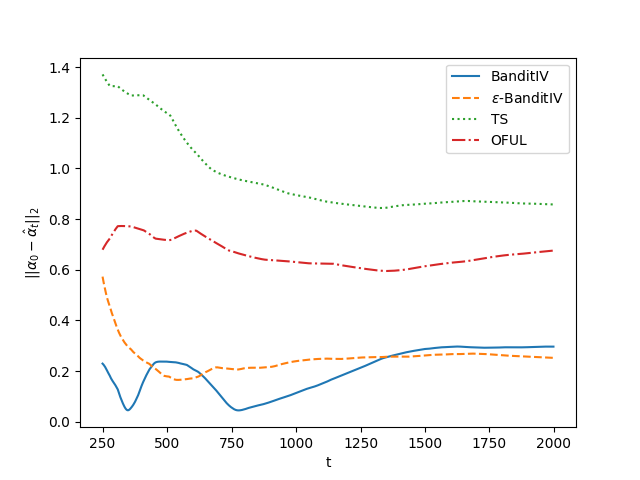}
        \caption{Estimation bias} %  when $\epsilon=0.0005$
        \label{fig:rtb_bias02_}
    \end{subfigure}
    \hfill
       \begin{subfigure}{0.3\textwidth}
        \centering
        \includegraphics[scale=0.3]{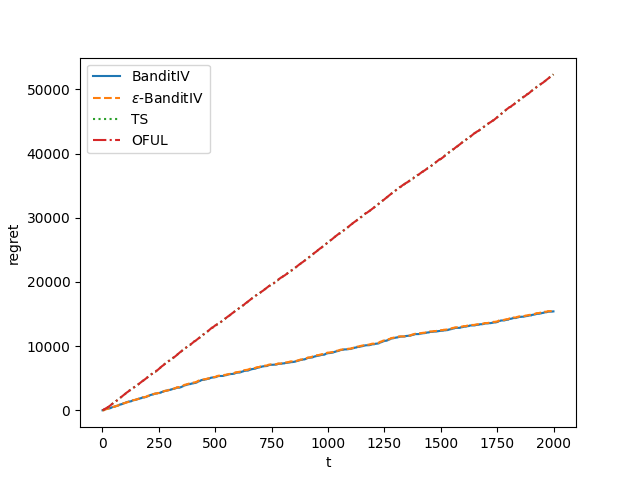}
        \caption{Regret variations} %  when $\epsilon=0.0005$
        \label{fig:rtb021_}
    \end{subfigure}
    \hfill
     \begin{subfigure}{0.3\textwidth}
        \centering
        \includegraphics[scale=0.3]{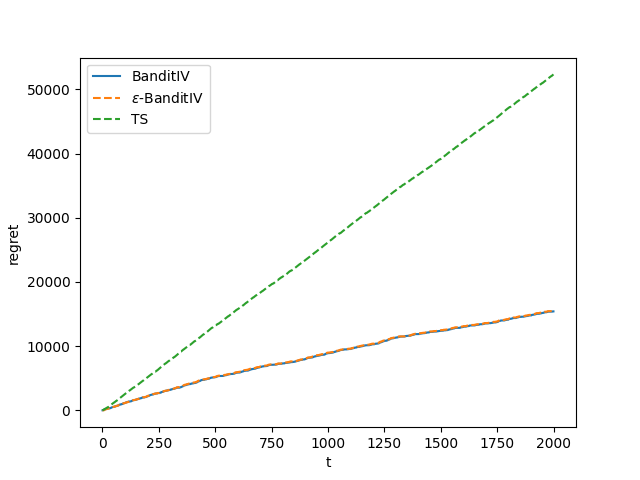}
        \caption{Regret variations} %  when $\epsilon=0.0005$
        \label{fig:rtb022_}
    \end{subfigure}

    \caption{Simulation results on RTB data with endogeneity: $\rho=0.5$}
    \label{fig:rtb1_}
\end{figure}

\section{Conclusion}
In this paper, we study the endogeneity problem in online decision-making settings where we formulate the decision-making process as a contextual linear bandit model. We discuss how many economic settings can be characterized by endogeneity and extant methods can lead to estimation bias and sub-optimal decisions. To correct the bias and optimize online decisions, we propose the $\epsilon$-\textit{BanditIV} algorithm by utilizing both existing linear bandit algorithms and instrumental variables. We present the theoretical properties of our algorithm. We first show an upper bound for the total expected regret, and then show the consistency, asymptotic normality of the estimator in the algorithm. On the applied side, we conduct extensive simulations on synthetic data. We find that the $\epsilon$-\textit{BanditIV} algorithm outperforms several benchmark linear bandit algorithms when the endogeneity problem occurs. Finally, we demonstrate by using data from the real-time bidding (RTB) system to demonstrate how $\epsilon$-\textit{BanditIV} can be effective in estimating the causal impact of advertising while maintaining close to oracle regret.

% \section*{References}
% \bibliographystyle{plainnat}
\bibliography{ref.bib}
% \printbibliography
% %%%%%%%%%%%%%%%%%%%%%%%%%%%%%%%%%%%%%%%%%%%%%%%%%%%%%%%%%%%%

\section*{Appendix}

\noindent We provide proofs for main results in Appendix A and auxiliary lemmas in Appendix B.

% Optionally include extra information (complete proofs, additional experiments and plots) in the appendix.
% This section will often be part of the supplemental material.
\subsection*{Appendix A. Proof of Main Results} \label{sec:appendixa}
\addcontentsline{toc}{section}{Appendix A}

\begin{repeattheorem}[\ref{thm: total_regret}]
    % \label{thm: total_regret}
The expected cumulative regret of the $\epsilon$-BanditIV at time T, with probability at least $1-\delta$, is upper-bounded by
\begin{align*}
    R_T &\le B_T \sqrt{2Td \log(\frac{T+d}{d})} + (\frac{2}{\gamma_x}+\|\beta_0\|_2)G_T\sqrt{2Tk \log(\frac{T+k}{k})} + 2\epsilon_0 T L_y
\end{align*}
where 
\begin{align*}
    &B_T\\ &=\sqrt{\gamma_x} ||\beta_0||_2+\sqrt{2\log(\frac{4T}{\delta})+  d \log\left((T \dfrac{k}{d}+ 2T^2  k \log (\frac{4d^2T}{\delta}))L_z^2||\Gamma_0||_F^2+T L_z  \sqrt{2\log (\frac{4d^2T}{\delta}) } ||\Gamma_0||_1\right) }
\end{align*},
$G_T=\sqrt{\gamma_z} \max_{i\in \{1,\cdots ,d\}}||\Gamma_0^{(i)}||_2+\sqrt{2\log(\frac{2Td}{\delta})+k \log\left(TL_z^2+k\gamma_z \right)}$.
\end{repeattheorem}

\proof{}
This regret can be decomposed into
\begin{align*}
    R_T
    & = \sum_{t=1}^T(1-\epsilon_t)\mathbb{E}_{u_t}[\langle x_t^*, \beta_0 \rangle - \langle x_t, \beta_0 \rangle] \one[x_t = \Gamma_0'z_t]
        + \sum_{t=1}^T \epsilon_t\mathbb{E}_{u_t}[\langle x_t^*, \beta_0 \rangle - \langle x_t, \beta_0 \rangle] \one[x_t \neq \Gamma_0'z_t]\\
    & \leq \sum_{t=1}^T\langle \Gamma'_0z_t^*, \beta_0 \rangle - \langle \Gamma'_0z_t, \beta_0 \rangle
        + 2\epsilon_0 T L_y
\end{align*}
Next we will upper bound the first term.
Let $(\Tilde{\Gamma}_t,\Tilde{\beta}_t)=\arg \max_{\Gamma \in G_{t-1}, \beta\in C_{t-1}} \langle \Gamma' z_t, \beta \rangle $. By using the \textit{two-stage} optimism we can decompose our regret as follows, 
\begin{align*}
    & \sum_{t=1}^T\langle \Gamma'_0z_t^*, \beta_0 \rangle - \langle \Gamma'_0z_t, \beta_0 \rangle  \\
    &\le \sum_{t=1}^T \max_{\Gamma \in  \calC_{2,t-1}} \max_{\beta\in  \calC_{2,t-1}}\langle \Gamma' z_t^*, \beta \rangle  - \langle \Gamma'_0z_t, \beta_0 \rangle\\
    &\le \sum_{t=1}^T \langle \Tilde{\Gamma}_t' z_t, \Tilde{\beta}_t \rangle  - \langle \Gamma'_0 z_t, \beta_0 \rangle\\
    % &= \underbrace{\sum_{t=1}^T \langle (\Tilde{\Gamma}_t-{\Gamma}_0)'z_t, \Tilde{\beta}_t \rangle}_\text{est error of $\Gamma_0$} 
    %     + \underbrace{\sum_{t=1}^T \langle \Gamma'_0 z_t, \Tilde{\beta}_t - \beta_0 \rangle }_\text{est error of $\beta_0$} 
    %     \label{eq:regret_2parts}
    & = \sum_{t=1}^T \langle (\Tilde{\Gamma}_t-{\Gamma}_0)'z_t, (\Tilde{\beta}_t-\beta_0) + \beta_0 \rangle + \langle  \hat{x}_t + (\Gamma_0-\hat{\Gamma}_t)'z_t, \Tilde{\beta}_t - \beta_0 \rangle\\
    & \leq \sum_{t=1}^T ||\hat{x}_t ||_{W_{t-1}^{-1}}\|\Tilde{\beta}_t - \beta_0 \|_{W_{t-1}}+ \sum_{t=1}^T ||(\Gamma_0-\hat{\Gamma}_t)'z_t ||_2 \| \Tilde{\beta}_t - \beta_0 \|_2 \\
    & \quad + \|\beta_0 \|_2 \sum_{t=1}^T \|(\Tilde{\Gamma}_t-{\Gamma}_0)'z_t\|_2 + \sum_{t=1}^T \|(\Tilde{\Gamma}_t-{\Gamma}_0)'z_t\|_2 \|\Tilde{\beta}_t - \beta_0\|_2\\
    & = \sum_{t=1}^T ||\hat{x}_t ||_{W_{t-1}^{-1}}\|\Tilde{\beta}_t - \beta_0 \|_{W_{t-1}} 
    + \|\beta_0 \|_2 \sum_{t=1}^T \|(\Tilde{\Gamma}_t-{\Gamma}_0)'z_t\|_2 + 2\sum_{t=1}^T ||(\Gamma_0-\hat{\Gamma}_t)'z_t ||_2 \| \Tilde{\beta}_t - \beta_0 \|_2 
\end{align*}
where the first inequality comes from the Lemma~\ref{prop:Bt},~\ref{prop:Gt} that $\Gamma_0 \in  \calC_{2,t}, \beta_0 \in  \calC_{1,t}$ for all $t$,  the second inequality comes from the definition of $z_t$, the second equality comes from the definition of $\hat{x}_t$  and some careful decomposition, and the last inequality comes from cauchy-swartz inequality. 

Now our goal is to upper bounded these three terms, which mainly comes form the estimation  error of $\Gamma_0$ in the first stage and the estimation error of $\beta_0$ in the second stage. For the first term, by applying Cauchy-Schwartz inequality and the standard elliptical potential lemma in \cite{carpentier2020elliptical}, we have
\begin{align*}
    \sum_{t=1}^T ||\hat{x}_t ||_{W_{t-1}^{-1}}\|\Tilde{\beta}_t - \beta_0 \|_{W_{t-1}}
    % & \leq \max_t \|\Tilde{\beta}_t - \beta_0 \|_{W_{t-1}} \sqrt{T \sum_{t=1}^T ||\hat{x}_t ||_{W_{t-1}^{-1}}^2} \\\
    % & \leq \max_t \|\Tilde{\beta}_t - \beta_0 \|_{W_{t-1}} \sqrt{T \yf{to be added}}
    & \leq \max_t \|\Tilde{\beta}_t - \beta_0 \|_{W_{t-1}}  \sum_{t=1}^T ||\hat{x}_t ||_{W_{t-1}^{-1}}\\
    &\leq \max_t \|\Tilde{\beta}_t - \beta_0 \|_{W_{t-1}} \sqrt{2Td \log(\frac{T+d}{d})}
\end{align*}

For the second term, by using similar arguments, we have that
\begin{align*}
    \|\beta_0 \|_2 \sum_{t=1}^T \|(\Tilde{\Gamma}_t-{\Gamma}_0)'z_t\|_2
    & \leq  \|\beta_0 \|_2 \sum_{t=1}^T \sum_{i=1}^d \mid (\Tilde{\Gamma}_t^{(i)}-{\Gamma}_0^{(i)})'z_t \mid  \\
    & \leq  \|\beta_0 \|_2 \sum_{t=1}^T \sum_{i=1}^d  \|\Tilde{\Gamma}_t^{(i)}-{\Gamma}_0^{(i)})\|_{U_{t-1}} \|z_t\|_{U_{t-1}^{-1}}   \\
    & \leq  \|\beta_0 \|_2 d \max_{t \in [T],i\in[d]} \|\Tilde{\Gamma}_t^{(i)}-{\Gamma}_0^{(i)})\|_{U_{t-1}} \sum_{t=1}^T \|z_t\|_{U_{t-1}^{-1}}  \\
    & \leq  \|\beta_0 \|_2 d \max_{t \in [T],i\in[d]} \|\Tilde{\Gamma}_t^{(i)}-{\Gamma}_0^{(i)})\|_{U_{t-1}} \sqrt{2Tk \log(\frac{T+k}{k})}
\end{align*}
where the last inequality again comes from standard ellipse potential lemma.

Finally, for the third term, by the definition of $W_t$, we have $\| \Tilde{\beta}_t - \beta_0 \|_2  \leq \frac{1}{\gamma}\|\Tilde{\beta}_t - \beta_0 \|_{W_{t-1}} $. Therefore, by repeating the proof in the second term, we have that,
\begin{align*}
    &2\sum_{t=1}^T ||(\Gamma_0-\hat{\Gamma}_t)'z_t ||_2 \| \Tilde{\beta}_t - \beta_0 \|_2 \\
    &\leq \frac{2}{\gamma} \max_t \|\Tilde{\beta}_t - \beta_0 \|_{W_{t-1}}  \max_{t \in [T],i\in[d]} \|\Tilde{\Gamma}_t^{(i)}-{\Gamma}_0^{(i)})\|_{U_{t-1}} \sqrt{2Tk \log(\frac{T+k}{k})}
\end{align*}

Therefore, by combing these three terms and the definition of $\calC_{1,t},\calC_{2,t}$, we complete the proof.

\endproof

\begin{repeatlemma}[\ref{prop:Bt}]

With high prob $1-\delta/2$, for all $t \in [T]$,
\begin{align*}
   ||\hat{\beta}_t-\beta_0||_{W_t}
   \le B_t
\end{align*}
.
% and\\
% $C_{\hat{x}}=L_z||\Gamma_0||_2+\sqrt{2d\log(\frac{1}{\epsilon})(1+{ L_z^2k\frac{\log (1+L_z^2T/\gamma_z)}{\log (1+\frac{1}{T})}})}$.
\end{repeatlemma}
\proof{}
At any fixed time $t$, we denote $\hat{x}_s^t = \Gamma_t' z_s$ for all $s \leq t$, so the $\hat{X}_t$ defined in the algorithm is a collection of all $\{\hat{x}_s^t\}_{s \leq t}$. For convenience, we drop the superscript $t$ here.
We also denote $\mathbf{e}_t \in \fR^t$ as a collection of $\{e_s\}_{s \leq t}$. 
Now we get the closed-form for $\hat{\beta}_t$ by ridge TSLS estimator as follows,
\begin{align*}
    \hat{\beta}_t
    &=W_t^{-1}Q_t\\
    &= W_t^{-1} \hat{X}_t' (X_t \beta_0 + \mathbf{e}_t)\\
    & =(X_t'P_{Z_t}X_t+\gamma_x I)^{-1}X_t'P_{Z_t}X_t\beta_0 + W_t^{-1} \hat{X}_t' \mathbf{e}_t \\
    % & =W_t^{-1}(W_t-\gamma_x I)\beta_0+W_t^{-1}S_t 
    & = \beta_0 - \gamma_x W_t^{-1}\beta_0 + W_t^{-1} \hat{X}_t' \mathbf{e}_t
\end{align*}
where we denote $Z_t(Z'_tZ_t)^{-1}Z_t$ as $P_{Z_t}$. The third equality comes from the definition of $\hat{W}_t$ and the upper bound of $\hat{X}_t$ in Lemma~
\ref{lem: some upper bounds}.
Therefore, we can write the estimation error of $\hat{\beta}_t$ compared to $\beta_0$ as 
\begin{align*}
    ||\hat{\beta}_t-\beta_0||_{W_t}
    &=||W_t^{-1} \hat{X}_t' \mathbf{e}_t-\gamma_x W_t^{-1}\beta_0||_{W_t}\\
    &=||\hat{X}_t' \mathbf{e}_t-\gamma_x \beta_0||_{W_t^{-1}}\\
    &\le ||\hat{X}_t' \mathbf{e}_t||_{W_t^{-1}}+\gamma_x ||\beta_0||_{W_t^{-1}}\\
    &\le ||\hat{X}_t' \mathbf{e}_t||_{W_t^{-1}}+\gamma_x ||\beta_0||_{(\gamma_x I)^{-1}}\\
    &= ||\hat{X}_t' \mathbf{e}_t||_{W_t^{-1}}+\sqrt{\gamma_x} ||\beta_0||_2
\end{align*}
Finally, by noticing that $\mathbf{e}_t$ is 1-subgaussian and the that $\exp (\langle q, \hat{X}_t' \mathbf{e}_t \rangle - \frac{||q||^2_{\hat{X}'_t\hat{X}'_t}}{2})$ for all $q \in \fR^d$ is a supermartingale, according to the Section 20.1 in \cite{lattimore2020bandit}, we have that, with probability $1-\delta/(4T)$, 
\begin{align*}
    ||\hat{X}_t' \mathbf{e}_t||_{W_t^{-1}} \le \sqrt{2\log(\frac{4T}{\delta})+\log (\frac{\det(W_t)}{\gamma_x^d})})
    .
\end{align*}
By combining the above result and the explicit calculation of $W_t$ detailed in Lemma~\ref{lem: some upper bounds}, we complete the proof. 
\endproof

\begin{repeatlemma}[\ref{prop:Gt}] 
With high prob $1-\delta/2$, for all $t \in [T]$ and all $i \in [d]$,
\begin{align*}
    ||\hat{\Gamma}_t^{(i)}-\Gamma_0^{(i)}||_{U_t}\le G_t
\end{align*}
.
\end{repeatlemma}

\proof{}
The proof steps are similar to the previous lemma. At any fixed time $t$ and dimension $i \in [d]$, we denote $\mathbf{u}_t^i \in \fR^t$ as a collection of $\{e_s\}_{s \leq t}$. We again get the closed-form of $\hat{\Gamma}_t$ as 
\begin{align*}
        \hat{\Gamma}_t=U_t^{-1}V_t=U_t^{-1}(U_t-\gamma_z I)\Gamma_0+U_t^{-1}Z_t'\mathbf{u}_t^i=\Gamma_0-\gamma_z U_t^{-1}\Gamma_0+U_t^{-1}Z_t'\mathbf{u}_t^i
\end{align*}
And therefore,
$
    ||\hat{\Gamma}_t^{(i)}-\Gamma_0^{(i)}||_{U_t}
    \le \|Z_t'\mathbf{u}_t^i\|_{U_t^{-1}}+\sqrt{\gamma_z} ||\Gamma_0^{(i)}||_2.
$

Finally, again by noticing that $\mathbf{u}_t^i$ is 1-subgaussian and the that $\exp (\langle q, \hat{X}_t' \mathbf{u}_t^i \rangle - \frac{||q||^2_{\hat{X}'_t\hat{X}'_t}}{2})$ for all $q \in \fR^d$ is a supermartingale, we have that, with probability $1-\delta/(2Td)$, 
\begin{align*}
    ||\hat{\Gamma}_t^{(i)}-\Gamma_0^{(i)}||_{U_t}
    \le \sqrt{\gamma_z} ||\Gamma_0^{(i)}||_2+\sqrt{2\log(\frac{2Td}{\delta})+\log (\frac{\det(U_t)}{\gamma_z^k})}
    \le G_t
\end{align*}
where the last inequality comes from the explicit calculation of $U_t$ detailed in Lemma~\ref{lem: some upper bounds} and the union bound over all $t \in [T]$.

\endproof

\begin{lemma}\label{lemma: prob_wv}
    Suppose $\{\mathcal{F}_t:t=1,\ldots,T\}$ is an increasing filtration of $\sigma$-fields. Let $\{W_t: t=1,\ldots, T\}$ be a sequence of variables such that $W_t$ is $\mathcal{F}_{t-1}$ measurable and $|W_t|\le L_w$ almost surely for all $t$. Let $\{v_t: t=1,\ldots, T\}$ be independent $\sigma_v$-subgaussian, and $v_t\perp \mathcal{F}_{t-1}$ for all $t$. Let $\mathcal{S}=\{s_1,\ldots,s_{|\mathcal{S}|}\}\subseteq \{1,\ldots, T\}$ be an index set where $|\mathcal{S}|$ is the number of elements in $\mathcal{S}$. Then for $\kappa>0$,
    \begin{equation*}
        P(\sum_{s\in \mathcal{S}}W_sv_s\ge \kappa)\le \exp(-\frac{\kappa^2}{2|\mathcal{S}|\sigma_v^2L_w^2})
    \end{equation*}
\end{lemma}

The proof of this lemma is provided in Lemma 1 of \cite{chen2021statistical}.

\begin{lemma}\label{lemma:dependent_ols_ineq}
    (Dependent OLS Tail Inequality). For the online decision making model, if all realizations of $z_t$ satisfy $||z_t||_{\infty}\le L_z$ for all $t$, and $\hat{\Sigma}=\frac{1}{t}\sum_{s=1}^tz_sz'_s$ has minimum eigenvalue $\lambda_{min}(\hat{\Sigma})>\lambda$ for some $\lambda>0$ almost surely. Then for any $\eta>0$,
    \begin{equation*}
        P(||\hat{\delta}_t-\delta_0||_1\le \eta)\ge 1-2k\exp{(-\frac{t\lambda^2\eta^2}{2k^2\sigma_v^2L_z^2})}
    \end{equation*}
        
\end{lemma}
\proof{}
    Based on the proofs provided by \cite{chen2021statistical} and \cite{bastani2020online}, we make some minor changes. The relation between eigenvalue and $l_2$ norm of symmetric matrix gives $||\hat{\Sigma}^{-1}||_2=\lambda_{max}(\hat{\Sigma}^{-1})=(\lambda_{min}(\hat{\Sigma}))^{-1}$. Therefore, 
    \begin{equation*}
        ||\hat{\delta}_t-\delta_0||_2=||\hat{\Sigma}^{-1}(\frac{1}{t}\sum_{s=1}^tz_sv_s)||_2\le \frac{1}{t}||\hat{\Sigma}^{-1}||_2||\sum_{s=1}^tz_sv_s||_2\le \frac{1}{t\lambda}||\sum_{s=1}^tz_sv_s||_2
    \end{equation*}
    Hence, we have
    \begin{align*}
         P(||\hat{\delta}_t-\delta_0||_2\le \eta)&\ge P(||\sum_{s=1}^tz_sv_s||_2\le t\lambda \eta)\\
         & \ge P(|\sum_{s=1}^tZ_{j,s}v_s|\le \frac{t\lambda \eta}{\sqrt{k}},\ldots, |\sum_{s=1}^tZ_{k,s}v_s|\le \frac{t\lambda \eta}{\sqrt{k}})\\
         &=1-P(\bigcup_{j=1}^k\{|\sum_{s=1}^tZ_{j,s}v_s|> \frac{t\lambda \eta}{\sqrt{k}}\})\\
         &\ge 1-\sum_{j=1}^kP(|\sum_{s=1}^tZ_{j,s}v_s|> \frac{t\lambda \eta}{\sqrt{k}})\\
    \end{align*}
    Because we know that $v_s$ in the above inequality are i.i.d. subgaussian and $v_s \perp z_s$, we can apply Lemma \ref{lemma: prob_wv} and have the following
    \begin{equation*}
        P(|\sum_{s=1}^tZ_{j,s}v_s|> \frac{t\lambda \eta}{\sqrt{k}})\le 2\exp(-\frac{t\lambda^2 \eta^2}{2k\sigma_v^2L_z^2})
    \end{equation*}
    Thus,
    \begin{equation*}
         P(||\hat{\delta}_t-\delta_0||_1\le \eta)\ge  P(||\hat{\delta}_t-\delta_0||_2\le \frac{\eta}{\sqrt{k}})\ge 1-2k\exp(-\frac{t\lambda^2 \eta^2}{2k^2\sigma_v^2L_z^2})
    \end{equation*} 
\endproof

\begin{lemma}\label{lemma:prob_bdamin}
    Let $\{z_t:t=1,\ldots,T\}$ be a sequence of i.i.d. $k$-dimension random vectors such that all realizations of $z_t$ satisfy $||z_t||_{\infty}\le L_z$ for all $t$. Denote $\hat{\Sigma}=\frac{1}{T}\sum_{t=1}^Tz_tz'_t$. If $\Sigma=\mathbb{E}[z_tz'_t]$ has minimum eigenvalue $\lambda_{min}(\Sigma)>\lambda$ for some $\lambda>0$, then
    \begin{equation*}
        P(\lambda_{min}(\hat{\Sigma})\le \frac{\lambda}{2})\le k \exp(-\frac{T \lambda}{8L_z^2})
    \end{equation*}
\end{lemma}
\proof{}
This proof is based on the proof of Lemma 3 in \cite{chen2021statistical} with minor changes. First, we have
\begin{equation*}
    \lambda_{max}(\frac{z_tz'_t}{T})=\max_{||a||_2=1}a'(\frac{z_tz'_t}{T})a=\frac{1}{T}\max_{||a||_2=1}(a'z_t)^2\le \frac{L_z^2}{T}
\end{equation*}
\begin{equation*}
   \mu_{min}\equiv \lambda_{min}(\mathbb{E}\hat{\Sigma})=\lambda_{min}(\frac{1}{T}\sum_{t=1}^t\mathbb{E}[z_tz'_t])=\lambda_{min}(\Sigma)>\lambda
\end{equation*}
Then, using the Matrix Chernoff bound, 
\begin{equation*}
     P(\lambda_{min}(\hat{\Sigma})\le \frac{\lambda}{2})\le P(\lambda_{min}(\hat{\Sigma})\le \frac{  \mu_{min}}{2})\le k \exp(-\frac{T\mu_{min}}{8L_z^2})\le k \exp(-\frac{T\lambda}{8L_z^2})
\end{equation*} 
\endproof

\begin{lemma}\label{lemma: prod_eigen}
    \begin{align*}
    ||\hat{\Gamma}(\hat{\beta} -\beta_{0})||_1 \ge \frac{\lambda_{min}(\hat{\Gamma})}{\sqrt{d}} ||\hat{\beta} -\beta_{0}||_1
\end{align*}
where $\lambda_{min}(\hat{\Gamma})$ is the smallest magnitude of a singular value of this matrix.
\end{lemma}
 
\proof{}
 By doing Singular Value Decomposition for $\hat{\Gamma}$, we have the following, where the second equality is due to that $U$ is orthonormal.
\begin{equation*}
     ||\hat{\Gamma}(\hat{\beta} -\beta_{0})||_2 =  ||U\Sigma V'(\hat{\beta} -\beta_{0})||_2 = ||\Sigma V'(\hat{\beta} -\beta_{0})||_2
\end{equation*}
% ||x||=(x'x)^(1/2) => ||Ux||=(x'U'Ux)^(1/2)=||x||

We assume $rank(\hat{\Gamma})= min\{k,d\}=d$. Using matrix form, we have 

\begin{align*}
    %\begin{gather}
     \Sigma V'(\hat{\beta} -\beta_{0})
     &=
      \begin{bmatrix}
       \sigma_1& 0& \cdots &0 \\
      0& \sigma_2& \cdots &0 \\
       \vdots & \vdots & \vdots &\vdots \\
       0& 0& \cdots & \sigma_d\\
       0& 0& \cdots & 0\\
        \vdots & \vdots & \vdots &\vdots \\
         0& 0& \cdots & 0\\
       \end{bmatrix}
         \begin{bmatrix}
       V^{(1)(1)}& V^{(2)(1)}& \cdots &V^{(d)(1)} \\
       V^{(1)(2)}& V^{(2)(2)}& \cdots &V^{(d)(2)} \\
       \vdots & \vdots & \vdots &\vdots \\
       V^{(1)(d)}& V^{(2)(d)}& \cdots &V^{(d)(d)}
       \end{bmatrix}
       \begin{bmatrix} \hat{\beta}^{(1)} -\beta_{0}^{(1)} \\ \hat{\beta}^{(2)} -\beta_{0}^{(2)} \\ \vdots\\ \hat{\beta}^{(d)} -\beta_{0}^{(d)} \end{bmatrix}\\
      & =
        \begin{bmatrix}  \sigma_1 \sum_{i=1}^d V^{(i)(1)}(\hat{\beta}^{(i)} -\beta_{0}^{(i)})  \\ \sigma_2 \sum_{i=1}^d V^{(i)(2)}(\hat{\beta}^{(i)} -\beta_{0}^{(i)}) \\ \vdots\\ \sigma_d \sum_{i=1}^d V^{(i)(d)}(\hat{\beta}^{(i)} -\beta_{0}^{(i)})\\
        0\\
        \vdots\\
        0
        \end{bmatrix}
   % \end{gather}
\end{align*}

Hence, 

\begin{align}\label{ineq: svd_beta}
       \nonumber || \Sigma V'(\hat{\beta} -\beta_{0})||_1\ge || \Sigma V'(\hat{\beta} -\beta_{0})||_2 &= \sqrt{\sum_{j=1}^d|\sigma_j \sum_{i=1}^d V^{(i)(j)}(\hat{\beta}^{(i)}-\beta_{0}^{(i)})|^2} \\ \nonumber
    &\ge \lambda_{min}(\hat{\Gamma})\sqrt{ \sum_{j=1}^d| \sum_{i=1}^d V^{(i)(j)}(\hat{\beta}^{(i)}-\beta_{0}^{(i)})|^2}\\ \nonumber
    &=  \lambda_{min}(\hat{\Gamma}) ||  V'(\hat{\beta} -\beta_{0})||_2 \\ 
    &= \lambda_{min}(\hat{\Gamma}) ||  (\hat{\beta} -\beta_{0})||_2
\end{align}
where the last equality is due to that $V$ is orthonormal.

Applying Cauchy–Schwarz inequality, we obtain
\begin{align}\label{ineq: dif_beta_nl1l2}
    ||(\hat{\beta} -\beta_{0})||_2\ge \frac{1}{\sqrt{d}} ||(\hat{\beta} -\beta_{0})||_1
\end{align}

Combing inequalities (\ref{ineq: svd_beta}) and (\ref{ineq: dif_beta_nl1l2}), we complete the proof. 
\endproof

\begin{lemma}\label{lemma: lambdamin_ineq}
  \begin{align}\label{ineq: sigma}
    \lambda_{min}(\hat{\Gamma})\ge \lambda_{min}(\Gamma_0)-\eta_2(kd)^{\frac{1}{2}}
\end{align}  
\end{lemma}

\proof{}
   On the event that $ ||\hat{\Gamma} - \Gamma_0||\le \eta_2$, we can rewrite the equation for each element in $\hat{\Gamma}$ using a newly defined matrix $A\in \mathbb{R}^{k\times d}: |A^{(i)(l)}|\le 1, \forall i \in i \in \{1,2,\ldots,d\}, \forall l\in \{1,2,\ldots,k\}$.
\begin{align}\label{eq: gamma_etaA}
     \hat{\Gamma}^{(i)(l)} =  \Gamma_0^{(i)(l)} + \eta_2A^{(i)(l)} 
\end{align}

We know that all singular values are non-negative. Here, we assume the singular values of $\Gamma_0$ are strictly positive. We can write the singular values of $\Gamma_0$ and $\hat{\Gamma}$ as two increasing sequences, $0 <\sigma_{\Gamma}^{(1)}\le\sigma_{\Gamma}^{(2)}\le \cdots \le \sigma_{\Gamma}^{(d)} $ and $ \sigma^{(1)}\le\sigma^{(2)}\le \cdots \le \sigma^{(d)} $, respectively.

Using the min-max principle for singular values, we have

\begin{align*}
    \sigma_{\Gamma}^{(1)} =\min_{S: dim(S)=1} \max_{\substack{a\in S \\ ||a||_2=1}}||\Gamma_0a||_2 
    = \min_{\substack{a\in \mathbb{R}^d \\ ||a||_2=1}}||\Gamma_0a||_2
\end{align*}

where the first equality is directly derived by applying the min-max theorem. The second equality is from the two facts. First, $a$ should be of an order $d\times 1$ to fit the order of $\Gamma_0$. Second, there are only two unique vectors fitting the constraint $||a||_2=1$ in the given space $S$ and they generate the same objective values $||\Gamma_0a||_2$.

Similarly, we can rewrite the smallest singular value of $\hat{\Gamma}$ as the following 
\begin{align}
    \sigma^{(1)} 
    = \min_{\substack{a\in \mathbb{R}^d \\ ||a||_2=1}}||\hat{\Gamma} a||_2
\end{align}

\begin{align*}
   \min_{\substack{a\in \mathbb{R}^d \\ ||a||_2=1}}||\hat{\Gamma} a||_2 &\ge \min_{\substack{a\in \mathbb{R}^d \\ ||a||_2=1}}||\Gamma_0 a||_2 - \eta_2||Aa||_2\\
   &\ge \min_{\substack{a\in \mathbb{R}^d \\ ||a||_2=1}}||\Gamma_0 a||_2 - \eta_2||A||_2||a||_2\\
    &= \min_{\substack{a\in \mathbb{R}^d \\ ||a||_2=1}}||\Gamma_0 a||_2 - \eta_2||A||_2\\
      &\ge \min_{\substack{a\in \mathbb{R}^d \\ ||a||_2=1}}||\Gamma_0 a||_2 - \eta_2(kd)^{\frac{1}{2}}\\
   &= \sigma_{\Gamma}^{(1)} - \eta_2(kd)^{\frac{1}{2}}
\end{align*}

where the first inequality is derived from the equation (\ref{eq: gamma_etaA}) and triangle inequality. The third equality is from the constraint that $||a||_2=1$. The fourth inequality is from the property of Frobenius norm, which we illustrate as the following

\begin{align*}
    ||A||_2\le ||A||_F = (\sum_{i=1}^d\sum_{l=1}^k |A^{(i)(l)}|^2)^{\frac{1}{2}}\le (\sum_{i=1}^d\sum_{l=1}^k 1)^{\frac{1}{2}}=(kd)^{\frac{1}{2}}
\end{align*} 

\endproof

\begin{repeatproposition}[\ref{prop: tail_ols}]
     In the online decision making model with the $\epsilon$-greedy policy, if Assumptions \ref{asm: l2norm} and \ref{asm: mineig} are satisfied, and $\epsilon_t$ is non-increasing, then for any $\eta_1, \eta_2>0$, any $i\in \{1,\ldots,d\}$,
    \begin{align*}
          P(||\hat{\delta}_t-\delta_0||_1\le \eta_1)&\ge 1-\exp(-\frac{t\epsilon_t}{8})-k\exp(-\frac{t\epsilon_t\lambda}{32L_z^2})-2k\exp(-\frac{t\epsilon_t^2\lambda^2\eta_1^2}{128k^2\sigma_v^2L_z^2})\\
          &+2k^2\exp(-\frac{t\epsilon_t^2\lambda^2\eta_1^2+4t\epsilon_t\lambda k^2\sigma_v^2}{128k^2\sigma_v^2L_z^2})
    \end{align*}
     \begin{align*}
          P(||\hat{\Gamma}_t^{(i)}-\Gamma_0^{(i)}||_1\le \eta_2)&\ge 1-\exp(-\frac{t\epsilon_t}{8})-k\exp(-\frac{t\epsilon_t\lambda}{32L_z^2})-2k\exp(-\frac{t\epsilon_t^2\lambda^2\eta_2^2}{128k^2\sigma_u^2L_z^2})\\
          &+2k^2\exp(-\frac{t\epsilon_t^2\lambda^2\eta_2^2+4t\epsilon_t\lambda k^2\sigma_u^2}{128k^2\sigma_u^2L_z^2})
    \end{align*}
\end{repeatproposition}

\proof{}\label{pf: prop_tail_ols}
    Following the proof of Proposition 3.1 in \cite{chen2021statistical}, we provide this proof adapted to our setting. Denote $\mathcal{S}_t=\{1,\ldots, t\}$ and define $\hat{\Sigma}(\mathcal{I})=|\mathcal{I}|^{-1}\sum_{s\in \mathcal{I}}z_sz'_s$ for any $\mathcal{I}\subseteq \mathcal{S}_t$, where $|\mathcal{I}|$ is the number of element in the set $\mathcal{I}$ and $|\mathcal{S}_t|=t$. We have
    \begin{equation*}
        \hat{\delta}_{t}-\delta_0=(\frac{1}{t}\sum_{s=1}^t z_sz'_s)^{-1}\frac{1}{t}\sum_{s=1}^t z_sv_s=\{\hat{\Sigma}(\mathcal{S}_t)\}^{-1}\frac{1}{t}\sum_{s=1}^t z_sv_s
    \end{equation*}
    Under $\epsilon$-greedy policy, we pull the estimated optimal arm with probability $\epsilon_s$ at each time point $s$. We denote the collection of time points up to time $t$ when the estimated optimal arm is chosen as $\mathcal{R}_t$. We will first bound the minimum eigenvalue of $\hat{\Sigma}(\mathcal{R}_t)$ and then use it to infer the bound of the minimum eigenvalue of $\hat{\Sigma}(\mathcal{S}_t)$.

    We denote the following event,
\begin{equation*}
    E:=\{\lambda_{min}(\hat{\Sigma}(\mathcal{R}_t))>\frac{\lambda}{4}\}
\end{equation*}
    Applying Lemma \ref{lemma:prob_bdamin}, if $\lambda_{min}(\Sigma)>\lambda$, then we have 
    \begin{equation*}
        P(E)\ge 1-k\exp(-\frac{|\mathcal{R}_t|\lambda}{16L_z^2})
    \end{equation*}
    Meanwhile, we know that 
    \begin{equation*}
        \hat{\Sigma}(\mathcal{S}_t)=\frac{|\mathcal{R}_t|}{|\mathcal{S}_t|}\hat{\Sigma}(\mathcal{R}_t)+\frac{|\mathcal{S}_t|-|\mathcal{R}_t|}{|\mathcal{S}_t|}\hat{\Sigma}(\mathcal{S}_t \backslash \mathcal{R}_t)
    \end{equation*}
    By Weyl's inequality, on event $E$,
    \begin{equation*}
        \lambda_{min}(\hat{\Sigma}(\mathcal{S}_t))\ge \lambda_{min}(\frac{|\mathcal{R}_t|}{|\mathcal{S}_t|}\hat{\Sigma}(\mathcal{R}_t)) + \lambda_{min}(\frac{|\mathcal{S}_t|-|\mathcal{R}_t|}{|\mathcal{S}_t|}\hat{\Sigma}(\mathcal{S}_t \backslash \mathcal{R}_t))\ge \frac{|\mathcal{R}_t|}{|\mathcal{S}_t|}\lambda_{min}(\hat{\Sigma}(\mathcal{R}_t))\ge \frac{\lambda|\mathcal{R}_t|}{4t}
    \end{equation*}
    Applying Lemma \ref{lemma:dependent_ols_ineq}, we have 
    \begin{equation*}
         P(||\hat{\delta}_t-\delta_0||_1\le \eta|E)\ge 1-2k\exp{(-\frac{|\mathcal{R}_t|^2\lambda^2\eta^2}{32tk^2\sigma_v^2L_z^2})}
    \end{equation*}
    Hence,
    \begin{align*}
         P(||\hat{\delta}_t-\delta_0||_1\le \eta)&\ge  P(||\hat{\delta}_t-\delta_0||_1\le \eta|E)P(E)\\ &\ge 1-2k\exp{(-\frac{|\mathcal{R}_t|^2\lambda^2\eta^2}{32tk^2\sigma_v^2L_z^2})}-k\exp(-\frac{|\mathcal{R}_t|\lambda}{16L_z^2})+2k^2\exp(-\frac{|\mathcal{R}_t|\lambda}{16L_z^2}-\frac{|\mathcal{R}_t|^2\lambda^2\eta^2}{32tk^2\sigma_v^2L_z^2})
    \end{align*}
    The checking step for that $|\mathcal{R}_t|$ is large enough is exactly the same as that in \cite{chen2021statistical}.
    We complete the proof by combining all the results above,
    \begin{align*}
          P(||\hat{\delta}_t-\delta_0||_1\le \eta)&\ge 1-\exp(-\frac{t\epsilon_t}{8})-k\exp(-\frac{t\epsilon_t\lambda}{32L_z^2})
          -2k\exp(-\frac{t\epsilon_t^2\lambda^2\eta^2}{128k^2\sigma_v^2L_z^2})\\
          &+2k^2\exp(-\frac{t\epsilon_t^2\lambda^2\eta^2+4t\epsilon_t\lambda k^2\sigma_v^2}{128k^2\sigma_v^2L_z^2})
    \end{align*}
    We can prove the inequality for $P(||\hat{\Gamma}_t^{(i)}-\Gamma_0^{(i)}||_1\le \eta_2)$ by similar steps. 
\endproof

\begin{repeatproposition}[\ref{lemma: online_beta}]
    In the online decision-making model with $\epsilon$-greedy policy, if the Assumptions \ref{asm: l2norm} and \ref{asm: mineig} are satisfied, and $\epsilon_t$ is non-increasing, then for any $\eta_1, \eta_2> 0$, $\eta_2\neq \frac{\lambda_{min}(\Gamma_0)}{(kd)^{\frac{1}{2}}} $,
    \begin{equation*}
P(||\hat{\beta}_t -\beta_{0}||_1 \le C_{\beta})\ge 1-(p_1+d p_2)
\end{equation*} 

where $C_{\beta}=\dfrac{\eta_1+  \eta_2||\beta_0||_1}{\lambda_{min}(\Gamma_0)d^{\frac{-1}{2}}-\eta_2(k)^{\frac{1}{2}}}$, 
\begin{align*}
    p_1= \exp(-\dfrac{t\epsilon_t}{8})+k\exp(-\dfrac{t\epsilon_t\lambda}{32L_z^2})+2k\exp(-\dfrac{t\epsilon_t^2\lambda^2\eta_1^2}{128k^2\sigma_v^2L_z^2})-2k^2\exp(-\dfrac{t\epsilon_t^2\lambda^2\eta_1^2+4t\epsilon_t\lambda k^2\sigma_v^2}{128k^2\sigma_v^2L_z^2}),
\end{align*}
\begin{align*}
    p_2=\exp(-\dfrac{t\epsilon_t}{8})+k\exp(-\dfrac{t\epsilon_t\lambda}{32L_z^2})+2k\exp(-\dfrac{t\epsilon_t^2\lambda^2\eta_1^2}{128k^2\sigma_u^2L_z^2})
-2k^2\exp(-\dfrac{t\epsilon_t^2\lambda^2\eta_1^2+4t\epsilon_t\lambda k^2\sigma_u^2}{128k^2\sigma_u^2L_z^2})
\end{align*}
\end{repeatproposition}

\proof{}
By Proposition \ref{prop: tail_ols}, we have $ ||\hat{\Gamma} - \Gamma_0||_1=\max_{1\le i\le d}||\hat{\Gamma}^{(i)} - \Gamma_0^{(i)}||_1\le \eta_2$ with a probability larger than $p_2^d$ and $ ||\hat{\delta} - \delta_0||_1\le \eta_1$ with a probability larger than $p_1$. On the events that $ ||\hat{\Gamma} - \Gamma_0||_1\le \eta_2$ and $ ||\hat{\delta} - \delta_0||_1\le \eta_1$, we can prove the upper bound of $||\hat{\beta} -\beta_{0}||_1$ as the following by utilizing triangle inequalities. 
\begin{align*}
||\hat{\Gamma}(\hat{\beta} -\beta_{0})||_1 -  ||(\hat{\Gamma}- \Gamma_{0})\beta_0||_1\le||\hat{\Gamma}(\hat{\beta} -\beta_{0}) + (\hat{\Gamma}- \Gamma_{0})\beta_0||_1=||\hat{\Gamma}\hat{\beta} - \Gamma_{0}\beta_{0}||_1 = ||\hat{\delta} - \delta_{0}||_1 \le \eta_1
\end{align*}
Hence, 
\begin{align}\label{ineq: prod_gm_beta}
    ||\hat{\Gamma}(\hat{\beta} -\beta_{0})||_1  \le  \eta_1 +  ||(\hat{\Gamma}- \Gamma_{0})\beta_0||_1 \le  \eta_1 +  ||\hat{\Gamma}- \Gamma_{0}||_1||\beta_0||_1 \le  \eta_1 +   \eta_2||\beta_0||_1
\end{align}
Combining inequality (\ref{ineq: prod_gm_beta}), Lemma \ref{lemma: prod_eigen} and Lemma \ref{lemma: lambdamin_ineq}, we complete the proof. 

\endproof

\textbf{Proof of "consistency of the variance estimator" in Theorem \ref{thm: asymnorm_banditiv}}
\proof{}
    We want to show that 
    \begin{align}\label{eq: var_p_converge}
         \sum_{s=1}^t \hat{v}_s^2(\hat{\Gamma}'_t \sum_{s=1}^t z_sz'_s \hat{\Gamma}_t)^{-1}\hat{\Gamma}'_t(\sum_{s=1}^t z_sz'_s)\hat{\Gamma}_t(\hat{\Gamma}'_t \sum_{s=1}^t z_sz'_s \hat{\Gamma}_t)^{-1} \overset{p}{\to} S
    \end{align}
    By rewriting the squared error term in (\ref{eq: var_p_converge}), we have
    \begin{align}\label{eq: rewrite_error_var}
          \frac{1}{t}\sum_{s=1}^t\hat{v}_s^2=  \frac{1}{t}\sum_{s=1}^t((\delta_0-\hat{\delta}_t)'z_s+v_s)^2=  \frac{1}{t}\sum_{s=1}^t((\delta_0-\hat{\delta}_t)'z_s)^2+ \frac{2}{t} \sum_{s=1}^t(\delta_0-\hat{\delta}_t)'z_sv_s+  \frac{1}{t}\sum_{s=1}^tv_s^2
    \end{align}
    Now we analyze the asymptotic properties for the three terms in (\ref{eq: rewrite_error_var}). Notice that by Proposition \ref{prop: tail_ols}, the first term in (\ref{eq: rewrite_error_var}) can be written as
    \begin{align*}
         (\delta_0-\hat{\delta}_t)' \frac{1}{t}\sum_{s=1}^t(z_sz'_s)(\delta_0-\hat{\delta}_t)\le L_z^2||\delta_0-\hat{\delta}_t||^2_1\overset{p}{\to} 0
    \end{align*}
    For the second term, by Proposition \ref{prop: tail_ols} and Lemma \ref{lm: consistency_zv}, 
    \begin{align*}
        (\delta_0-\hat{\delta}_t)'\frac{2}{t} \sum_{s=1}^tz_sv_s \overset{p}{\to} 0
    \end{align*}
    
    For the third term, we apply the weak law of large numbers and obtain
    \begin{align*}
        \frac{1}{t}\sum_{s=1}^tv_s^2 \overset{p}{\to} \mathbb{E}[v_s^2]
    \end{align*}
    
    Combining the results above and Proposition \ref{prop: tail_ols} with continuous mapping theorem, we complete the proof. 
\endproof

\subsection*{Appendix B. Auxiliary Lemmas} \label{sec:appendixb}
\addcontentsline{toc}{section}{Appendix B}

\begin{lemma}
\label{lem: some upper bounds}
For each $t$, we can explicitly upper bound the following terms,
\begin{align*}
    & W_t \leq X_t' Z_t' (Z_t'Z_t)^{-1}Z_t X_t, \\ & \log(\det(W_t))\\
    &\le  d \log\left(\frac{1}{d}\left((t k + 2t^2dk \log (\frac{4d^2T}{\delta}))L_z^2 ||\Gamma_0||_F^2+t L_z  \sqrt{2\log (\frac{4d^2T}{\delta}) } d||\Gamma_0||_1\right)\right) ,\text{w.p }\ge (1-\delta/(4T))\\
    & U_t \leq Z_t'Z_t + \gamma_zI, \quad  \log(\det(U_t)) \leq  k \log\left(tL_z^2+k\gamma_z\right)\\
    & \hat{X}_t \leq P_{Z_t} X_t 
\end{align*}     
\end{lemma}
\proof{}
For the first inequality,
\begin{align*}
    \hat{X}_t 
    = Z_t \hat{\Gamma}_t
    = Z_t (Z_t'Z_t + \gamma_z I)^{-1}Z_t' X_t
    \le  Z_t (Z_t'Z_t)^{-1}Z_t' X_t.
\end{align*}
Then for the second inequality, 
\begin{align*}
    W_t
    & = \hat{X}_t' \hat{X}_t\\
    & = (Z_t \hat{\Gamma}_t)'(Z_t \hat{\Gamma}_t)\\
    & = (Z_t (Z_t'Z_t + \gamma_z I)^{-1}Z_t' X_t)'(Z_t (Z_t'Z_t + \gamma_z I)^{-1}Z_t' X_t) \\
    & = X_t' Z_t (Z_t'Z_t + \gamma_z I)^{-1} Z_t' Z_t (Z_t'Z_t + \gamma_z I)^{-1}Z_t' X_t \\
    & \leq X_t' Z_t (Z_t'Z_t)^{-1}Z_t' X_t
\end{align*}
Recall that $P_{Z_t} = Z_t' (Z_t'Z_t)^{-1}Z_t $, we finish the upper bound of $W_t$.Then we have
\begin{align*}
    \log(\det(W_t)) 
    & \leq d \log\left(\frac{1}{d}\text{trace}(W_t)\right)\\
    & \leq d \log\left(\frac{1}{d}\text{trace}((Z_t'Z_t)^{-1}Z_t' X_t X_t' Z_t )\right)\\
    & = d \log\left(\frac{1}{d}\text{trace}((Z_t'Z_t)^{-1}Z_t' (Z_t\Gamma_0 + \boldsymbol{u}_t)(Z_t\Gamma_0 + \boldsymbol{u}_t)' Z_t )\right)\\
    & = d \log\left(\frac{1}{d}\left(\text{trace}(\Gamma_0\Gamma_0'Z_t'Z_t) + \text{trace}(\Gamma_0\Gamma_0'Z_t'\boldsymbol{u}_t \boldsymbol{u}_t' Z_t) + 2\text{trace}(Z_t'\boldsymbol{u}_t\Gamma_0') \right)\right) 
\end{align*}
where $u_t=X_t-Z_t\Gamma_0$, the first inequality is from the AM-GM inequality, the second inequality is from that $W_t$ is positive definite. We have assumed that each element of $u_t$ is 1-subgaussian and $u_t\in \mathbb{R}^{t\times d}$. 
\begin{align*}
   \text{trace}(\Gamma_0\Gamma_0'Z_t'\boldsymbol{u}_t \boldsymbol{u}_t' Z_t)&= \text{trace}(u'_tZ_t\Gamma_0\Gamma_0'Z_t'u_t)
    =||u'_tZ_t\Gamma_0||_F^2
\end{align*}
\begin{align*}
   \text{trace}(\Gamma_0\Gamma_0'Z_t'Z_t)&= \text{trace}(Z_t\Gamma_0\Gamma_0'Z_t')
    =||Z_t\Gamma_0||_F^2
\end{align*}
\begin{align*}
    \text{trace}(Z_t'\boldsymbol{u}_t\Gamma_0')=\text{trace}(\boldsymbol{u}_t\Gamma_0'Z_t')=\sum_{i=1}^d\sum_{j=1}^{t}u_t^{(i)(j)}\Tilde{X}_t^{(i)(j)}=\sum_{i=1}^d \Tilde{u}_t^{(i)(i)}
\end{align*}
% $\boldsymbol{u}_t\boldsymbol{u}_t' < 2I$. 

% We denote the $\tau^{th}$ row of $u_t$ as $(u_t^{[\tau]})'$, the $i^{th}$ element in the $\tau^{th}$ row of $u_t$ as $u_t^{[\tau](i)}$
We denote the matrix $u'_tZ_t\Gamma_0$ as $\Tilde{u}_t$ and $Z_t\Gamma_0$ as $\Tilde{X}_t$. We have that $|Z_t^{(i)(j)}|\le L_z$ for $\forall i\in \{1,\cdots, k\}, j\in \{1,\cdots, t\}$ by the Assumption \ref{asm: l2norm}. Therefore, we can derive the following, by the Cauchy–Schwarz inequality, 
\begin{align*}
    \sum_{\tau=1}^t (\Tilde{X}_t^{(j)(\tau)})^2\le \sum_{\tau=1}^t  (L_z \sum_{h=1}^k \Gamma_0^{(j)(h)})^2=tL_z^2 (\sum_{h=1}^k \Gamma_0^{(j)(h)})^2 
\end{align*}
\begin{align*}
     ||Z_t\Gamma_0||_F^2=\sum_{j=1}^d \sum_{\tau=1}^t (\Tilde{X}_t^{(j)(\tau)})^2\le tL_z^2 \sum_{j=1}^d(\sum_{h=1}^k \Gamma_0^{(j)(h)})^2 \le tL_z^2 k\sum_{j=1}^d\sum_{h=1}^k (\Gamma_0^{(j)(h)})^2 = tL_z^2 k ||\Gamma_0||_F^2
\end{align*}

Hence, by the Hoeffding's inequality, with a probability not less than $1-\delta/(4d^2T)$, 
\begin{align*}
    \Tilde{u}_t^{(i)(j)}=\sum_{\tau=1}^t u_t^{(i)(\tau)}\Tilde{X}_t^{(j)(\tau)} < t\sqrt{2\log (\frac{4d^2T}{\delta})L_z^2 (\sum_{h=1}^k \Gamma_0^{(j)(h)})^2 }.
\end{align*}

By the Boole’s inequality, we know that with a probability not less than $1-\delta/(4T)$, 
\begin{align*}
    ||u'_tZ_t\Gamma_0||_F^2 &<\sum_{j=1}^d 2t^2d \log (\frac{4d^2T}{\delta})L_z^2 (\sum_{h=1}^k \Gamma_0^{(j)(h)})^2 \\
    &\le \sum_{j=1}^d 2t^2d \log (\frac{4d^2T}{\delta})L_z^2 k\sum_{h=1}^k (\Gamma_0^{(j)(h)})^2 \\
    &=  2t^2dk \log (\frac{4d^2T}{\delta})L_z^2 ||\Gamma_0||_F^2.
\end{align*}
,
\begin{align*}
    \sum_{i=1}^d \Tilde{u}_t^{(i)(i)} < t L_z  \sqrt{2\log (\frac{4d^2T}{\delta}) } \sum_{i=1}^d\sum_{h=1}^k |\Gamma_0^{(i)(h)}| \le t L_z  \sqrt{2\log (\frac{4d^2T}{\delta}) } d||\Gamma_0||_1
\end{align*}
where the last line of inequality is from the triangular inequality.

Then $\log(\det(W_t))$ is upper bounded by the following, with a probability $1-\delta/(4T)$,
\begin{align*}
    d \log\left(\frac{1}{d}\left(tL_z^2 k ||\Gamma_0||_F^2+ 2t^2dk \log (\frac{4d^2T}{\delta})L_z^2 ||\Gamma_0||_F^2+t L_z  \sqrt{2\log (\frac{4d^2T}{\delta}) } d||\Gamma_0||_1\right)\right) 
    % \leq d \log\left(\frac{5TL_x^2}{d} \right)
\end{align*}
%
% Let $\delta=1/T$, we have 
% $$\log(\det(W_t))\le  d \log\left(\frac{1}{d}\left(tL_z^2 k ||\Gamma_0||_F^2+ 2t^2dk \log (T)L_z^2 ||\Gamma_0||_F^2+t L_z  \sqrt{2\log (T) } d||\Gamma_0||_1\right)\right)$$ with probability $(1-\frac{1}{T})^{d^2}$.
Similarly, we have 
\begin{align*}
    \log(\det(U_t)) 
    & \leq k \log\left(\frac{1}{k}\text{trace}(U_t)\right)\\
    &\le k \log\left(\frac{1}{k}\text{trace}(Z'_tZ_t+\gamma_zI)\right)\\
    & = k \log\left(\frac{1}{k}\text{trace}(Z'_tZ_t)+k\gamma_z\right)\\
    &= k \log\left(\frac{1}{k}||Z_t||_F^2+k\gamma_z\right)\\
    & \le k \log\left(tL_z^2+k\gamma_z\right)
\end{align*}

% TO CHECK
% \begin{align*}
%     \log(\det(U_t)) 
%     \leq k \log\left(\frac{5TL_y^2}{k} \right)
% \end{align*} 
% \Halmos
\endproof

%\iffalse
\begin{lemma}\label{lm:wagen}
    Consider a sequence of vectors $(x_t)_{t=1}^T, x_t \in \mathbb{R}^d$, and assume that $||x_t||_2 \le a$ for all $t$. Let $V_t = \lambda I + \sum_{s=1}^t x_s x'_s$ for some $\lambda > 0$. Then, we will have that $|| x_t ||_{V_{t-1}^{-1}} > b$ at most
\begin{align*}
d \log ( 1 + a^2 T/\lambda)/\log ( 1 + b)
\end{align*}
times. 
\end{lemma}
Lemma \ref{lm:wagen} is directly from lemma 6.2 in \cite{wagenmaker2021first}.

\begin{lemma}\label{lm:supermtg}
    $a_1, a_2,\cdots \in \mathbb{R}^d$ is $\mathcal{F}_t$ adapted and $\mathbb{E}[a_t|\mathcal{F}_{t-1}]=0$ and let $L_t\in \mathbb{R}^{d\times d}$ be a predictable sequence such that for any $\lambda \in \mathbb{R}^d$, 
\begin{align*}
    \mathbb{E}[\exp{(\langle \lambda, a_t \rangle)|\mathcal{F}_{t-1}}]\le \exp{(||\lambda||^2_{L_t}/2)}
\end{align*}
Let $H_t=\sum_{s=1}^ta_s$, $J_t=\sum_{s=1}^tL_t$.
\begin{align*}
    K_t(\lambda)=\exp{(\langle \lambda, H_t \rangle - ||\lambda||^2_{J_t}/2)}
\end{align*}
is a super martingale. Moreover, we have that
\begin{align*}
    \mathbb{P}(\exists t: ||H_t||_{(J_t+\gamma_x I)^{-1}} \ge \sqrt{2\log (1/\delta)+\log (\frac{det(J_t+\gamma_x I)}{\gamma_x^d})})\le \delta
\end{align*}

\end{lemma}

Lemma \ref{lm:supermtg} is from Theorem 2 in \cite{abbasi2011improved}
% from the lemma for vector-valued martingale in Kevin's lecture notes (Proposition 3 in Section 3.3.5).

\textbf{Proof of "each element of $v_{t,a}$ is $(||\beta_0||_2^2 +1)$-subgaussian"}
\label{pf: v_subgaussian}
\proof{}
We denote the $i^{th}$ element of $v_{t,a}$ as  $v_{t,a}^{(i)}$. To make the notation simpler, we omit the subscript ${t,a}$ in this proof.
     \begin{align*}
    \mathbb{E}[exp(\lambda v^{(i)})] &= \mathbb{E}[exp(\lambda (\sum_{l=1}^d u^{(i)(l)}\beta_0^{(l)}+e^{(i)}))]\\
    &= \prod_{l=1}^d \mathbb{E}[exp(\lambda u^{(i)(l)}\beta_0^{(l)})]\mathbb{E}[exp(\lambda e^{(i)})]\\
    &\le \prod_{l=1}^d \mathbb{E}[exp(\frac{\lambda^2 \beta_0^{(l)^2}}{2})]\mathbb{E}[exp(\frac{\lambda^2}{2})]\\
    &= exp(\frac{\lambda^2}{2}(\sum_{l=1}^d \beta_0^{(l)^2} +1))\\
    &= exp(\frac{\lambda^2}{2}(||\beta_0||_2^2 +1))
 \end{align*}
 where the second equality is due to the independency of distributions of $u^{(i)(l)}$ and $e^{(i)}$. The third inequality is because that $u^{(i)(l)}$ and $e^{(i)}$ are 1-subgaussian with mean zero. 
 \endproof

\textbf{Proof of "$\hat{\delta} = \hat{\Gamma} \hat{\beta}_{X,Y}$"}

\proof
    We want to prove the following equation
    \begin{align}\label{eq: gamma_beta_delta}
        \hat{\Gamma}\hat{\beta}_{X,Y}=\hat{\delta}
    \end{align}
    Multiplying $((Z\hat{\Gamma})'Z\hat{\Gamma})^{-1}(Z\hat{\Gamma})'Z$ to both sides of Equation (\ref{eq: gamma_beta_delta}), we can obtain the exact form of TSLS estimator which we defined before. Therefore, we can complete the proof. 

\endproof

\begin{lemma}\label{lm: consistency_zv}
    Suppose $\{\mathcal{F}_t: t=1,\ldots,T\}$ is an increasing filtration of $\sigma$-fields. Let $\{\tilde{W}_t: t=1,\ldots,T\}$ be a sequence of random variables such that $\tilde{W}_t$ is $\mathcal{F}_{t-1}$ measurable and $|\tilde{W}_t|\le L_{\tilde{w}}$ almost surely for all $t$. Let $\{\tilde{e}_t:t=1,\ldots,T\}$ be independent $\sigma_{\tilde{e}}$-subgaussian, and $\tilde{e}_t\perp \mathcal{F}_{t-1}$ for all $t$. Let $\tilde{S}=\{s_1,\ldots,s_{|\tilde{S}|}\}\subseteq \{1,\ldots,T\}$ be an index set where $|\tilde{S}|$ is the number of elements in $\tilde{S}$. Then for $\kappa >0$,
    \begin{align}
        P(\sum_{s\in \tilde{S}} \tilde{W}_s\tilde{e}_s)\le \exp \{-\frac{\kappa^2}{2|\tilde{S}|\sigma_{\tilde{e}}^2}L_{\tilde{w}}^2\}
    \end{align}
    
\end{lemma}
Lemma \ref{lm: consistency_zv} is from Lemma 1 in \cite{chen2021statistical}.

\end{document}